\documentclass[]{article} 
\usepackage[utf8]{inputenc}
\usepackage{amsmath}
\usepackage{graphicx}
\usepackage[a4paper, total={6in, 8in}]{geometry}
\usepackage{listings}
\usepackage{float}
\usepackage{placeins}
\usepackage[nottoc,numbib]{tocbibind}
\usepackage{xcolor}
\usepackage{hyperref}
\hypersetup{colorlinks,linkcolor={blue},citecolor={blue},urlcolor={blue}}
\usepackage[]{lineno}
\usepackage{bm}
\usepackage{url}
\usepackage[affil-it]{authblk}

\title{3D PIC Simulations on Hall Thruster Electron Drift
Instability:
Influence of Magnetic Field on Electron Transport}

\author[1,*]{Yinjian Zhao}
\author[1]{Kunpeng Zhong}

\affil[1]{School of Energy Science and Engineering, Harbin Institute of Technology, Harbin 150001, People’s Republic of China}
\affil[*]{Corresponding author: Yinjian Zhao,
zhaoyinjian@hit.edu.cn}

\date{\today}

\begin{document}

% \twocolumn[
%   \begin{@twocolumnfalse}
    \maketitle
    \begin{abstract}

Three-dimensional particle-in-cell simulations are employed to investigate
electron transport characteristics in Hall thrusters,
with particular focus on how magnetic field configuration affects
the electron transport due to
electron drift instabilities.
Comparing analytic and realistic magnetic field models reveals
significant differences in electron transport patterns,
where radial variations in field strength lead to asymmetric transport
enhanced in low-field regions.
The derived effective electron mobility shows agreement with direct simulation diagnoses,
and the obtained two-dimensional transport profiles provide a
foundation for developing more accurate reduced-dimensional models.

    \end{abstract}
%   \end{@twocolumnfalse}
% ]

\section{Introduction}

Hall thrusters are among the most promising electric propulsion technologies for space missions.
However, optimizing Hall thrusters requires a fundamental understanding of the physics behind the electron drift instabilities (EDI). The EDI is viewed as a kinetic instability that forms due to the large electron drift velocity (relative to the ions) in the azimuthal $E\times B$ direction, and represents a coupling between electron Bernstein modes and ion acoustic waves \cite{Ducrocq,Tsikata,Lafleur3}. To date, this instability has been observed both in experiments \cite{Tsikata2} and simulations.
Although many studies have utilized two-dimensional particle-in-cell (PIC) simulations, such as those in the azimuthal-axial \cite{Charoy2019} and azimuthal-radial \cite{Villafana2021} configurations, 
such approaches have certain limitations and may
be insufficient for capturing the full complexity
of the physics\cite{Boeuf}.
Thus, the implementation of three-dimensional PIC simulations becomes essential.

In PIC simulations of Hall thruster EDI, conventional acceleration methods inevitably introduce substantial physical compromises.
Early 3D PIC implementations, such as those by Hirakawa and Arakawa in the 1990s \cite{1996Numerical}, used reduced ion-to-electron mass ratios and modified vacuum permittivity to lower computational costs, that would alter the evolution of instabilities.
Later geometric scaling approaches, exemplified by Minelli and Taccogna’s 2018 downscaled SPT-100 model \cite{Minelli,Taccogna}, achieved qualitative agreement with experiments, but inherently amplified plasma-wall interactions due to increased surface-to-volume ratios.
Such scaling strategies risk distorting EDI growth rates and saturation dynamics by concurrently modifying electron cyclotron frequencies and ion convection timescales.
A paradigm shift emerged in 2023, when Villafana et al. \cite{Villafana2023} demonstrated the first full-scale 3D PIC simulation free from acceleration methods. Their centimeter-scale model employed 13.5 million unstructured tetrahedral cells spanning the discharge channels and near-plume regions, requiring 1.4 million CPU core-hours on 1,920 processors to simulate 10 microseconds.
In 2024, A. Denig and K. Hara applied
3D PIC simulations in a periodic box on EDI,
focusing on a long-wavelength mode occurred
after the instability saturation
\cite{IEPC-2024-852}.
In 2025, Denig and Hara further elucidated the critical roles of radial boundary conditions and ion temperature in governing the development of EDI \cite{IEPC-2025-492}.

The influence of the magnetic field was studied before
in a number of references.
For example,
Lafleur et al. (2016) developed a 1D simulation model to investigate the effects of a uniform radial magnetic field configuration on azimuthal instability development.
They found that an increased magnetic field or a decreased electron density will reduce the effect of the instability on the macroscopic plasma transport, and if the magnetic field is increased too much, the instability becomes too weak and the associated electron–ion friction can no longer enhance the electron cross-field transport \cite{Lafleur1}.
Reza et al. (2023) used an axial-azimuthal quasi-2D simulation framework to analyze the impact of radial magnetic field configurations on the evolution of azimuthal instabilities, and they noticed that the evolution and the characteristics of the azimuthal instabilities were majorly affected by the radial B-field gradient \cite{Reza_2023}.
And they also employed a quasi-2D simulation model incorporating partially real radial magnetic field configurations, and demonstrated that the gradients in the magnetic field configuration affect the spectrum of the azimuthal instabilities, which consequently changes the dominant mechanism behind the electron axial mobility \cite{R2}.
Currently, no 3D PIC simulations with real
magnetic fields exist for Hall thruster azimuthal instabilities, which defines the goal of our work.

Over recent years, our research group has pursued a series of 3D PIC studies focused on EDI in Hall thrusters. Our initial investigation \cite{Xie_2024} examined how plasma initialization influences simulation convergence,
an aspect often negligible in lower-dimensional studies but critical for avoiding days or weeks of unnecessary computation in large-scale 3D simulations. This was followed by the development of the PMSL code, used to analyze how plume region configuration affects EDI dynamics \cite{10.1063/5.0253669}. In both studies, simplified analytic models for ionization and magnetic fields were adopted, consistent with the approaches used in established benchmarks \cite{Villafana2021, Charoy2019} and in Villafana's 3D work \cite{Villafana2023}.

Building on this foundation, our most recent work introduced a more physically comprehensive PIC model, incorporating a Monte Carlo collision treatment for ionization coupled with a self-consistent fluid solver for neutral gas density. This framework was applied to specifically investigate the influence of realistic magnetic fields including radial and axial components on EDI behavior \cite{zhong20253dpicstudymagnetic}.

In our previous work \cite{zhong20253dpicstudymagnetic}, we presented preliminary results for three magnetic field configurations: strong realistic, weak realistic, and analytic magnets. However, those simulations were limited by a relatively small macro-particle population.
This constraint resulted in significant numerical noise, preventing more detailed physical analysis. To address this, the present study performs follow-up simulations with an increase in macro-particle number. This enhancement enables a comprehensive analysis of both transient and time-averaged field properties, including electron density ($n_e$), azimuthal electric field ($E_y$), their correlation $n_e E_y$, electron mobility ($\mu_e$), and electron/ion temperatures and average velocities,
that were also preliminarily presented in a recent conference
paper \cite{IEPC-2025-063}.

In this paper, we continue analyzing the dataset generated by the 3D PIC simulations presented in \cite{IEPC-2025-063}, with an additional and a specific focus on the influence of the magnetic field on electron transport. The time and space averaged
correlation term between the electron density and the electric field
$\left< n_e E_y \right>$, the effective mobility (excluding collisions) $\mu_\textrm{eff}'$, and the PIC mobility $\mu_{ez}$ from three simulation cases under different magnetic fields are presented and discussed for various time periods during the low-frequency ionization breathing mode. The simulation setup is detailed in Section \ref{sec:setup}, results are presented and discussed in Section \ref{sec:results}, and conclusions are summarized in Section \ref{sec:conclusion}.

\begin{figure}[H]
\centering
\includegraphics[width=0.4\textwidth]{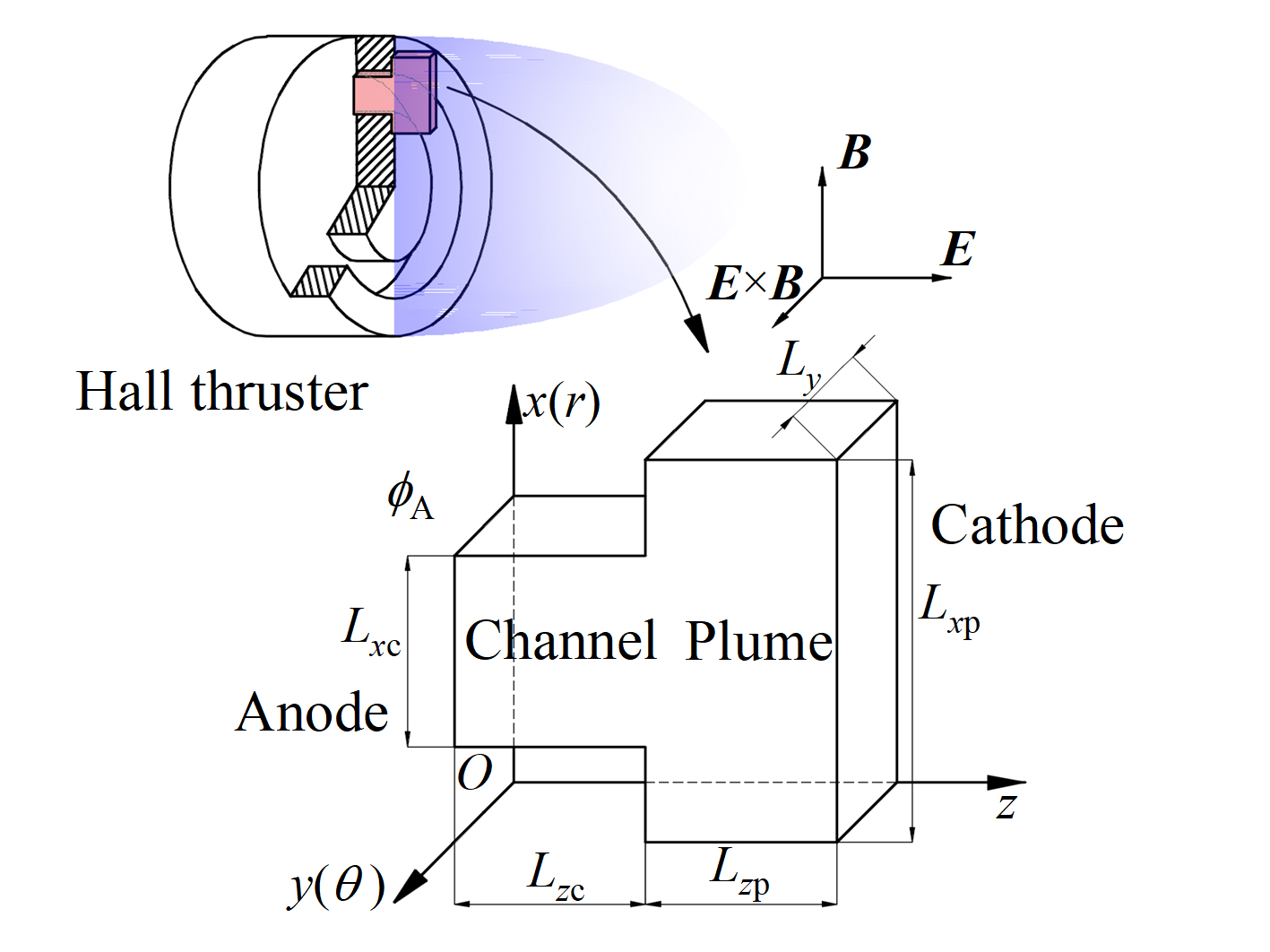}
\caption{
Illustration of the
simulation domain.
}
\label{fig:domain}
\end{figure}

\section{Simulation Setup}
\label{sec:setup}

The simulation domain and numerical setup maintain the same configuration as our previous studies \cite{zhong20253dpicstudymagnetic, IEPC-2025-063}, but are briefly summarized below for completeness. As shown in Fig.~\ref{fig:domain}, the computational domain comprises a discharge channel and a plume region. The coordinate system is defined with the $x$-, $y$-, and $z$-axes corresponding to the radial ($r$), azimuthal ($\theta$), and axial directions, respectively.
The anode is positioned at $z = 0$ with a fixed potential $\phi_A = 200$ V. Other boundaries in the $x$ and $z$ directions are treated as either wall boundaries or vacuum boundaries, both maintained at zero potential. Periodic boundary conditions are applied to the $y$-direction surfaces. Particles reaching the anode, cathode, walls, or vacuum boundaries are removed from the simulation.
Electron-ion pairs generated through ionization are modeled using the classical Monte Carlo Collision (MCC) method \cite{VAHEDI1995179}, with only ionization collisions considered. To maintain global charge neutrality and provide ionization sources, free electrons are injected through the cathode boundary. Ions are modeled with xenon mass.

\begin{figure}[ht]
\centering
(a)
\includegraphics[width=0.4\textwidth]{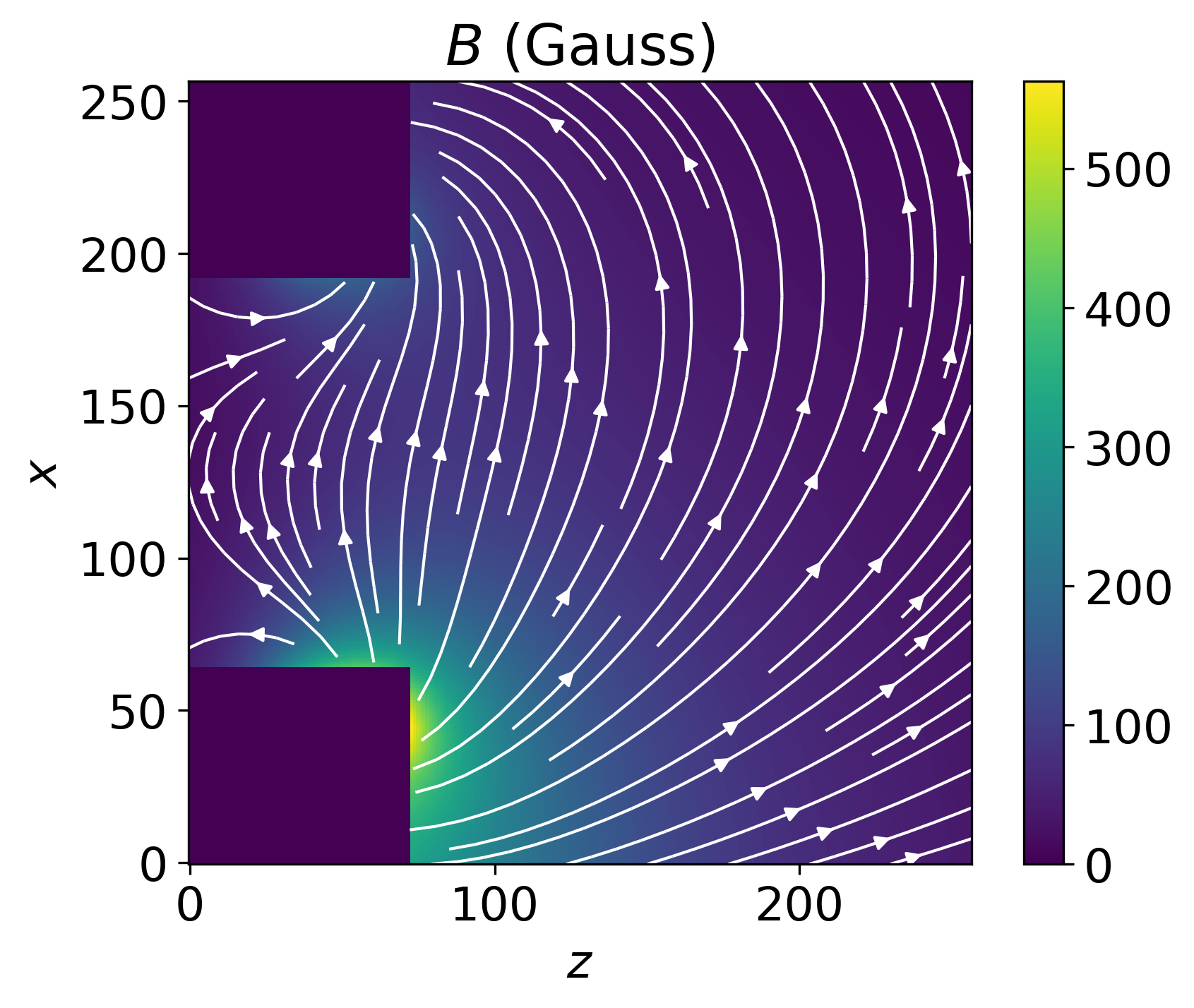}
(b)
\includegraphics[width=0.4\textwidth]{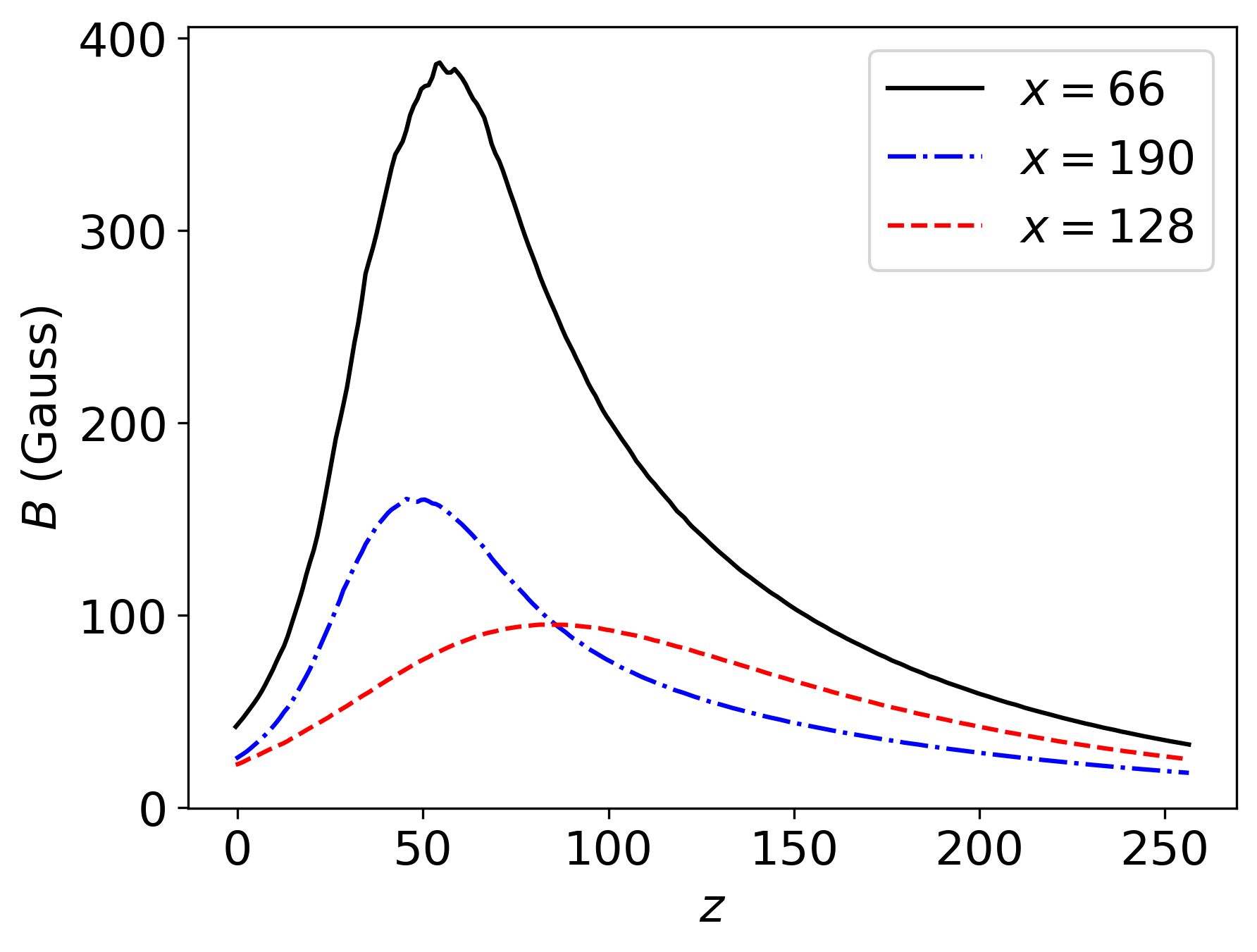}
\caption{
Magnetic field configuration
(a)
and 1D magnetic field strength at
three radial positions along $z$ (b)
of the Weak-B case (and the Strong-B case
with doubled field strength).
The $z$ and $x$ values represent grid indices.
}
\label{fig:B}
\end{figure}

\begin{figure}[ht]
\centering
\includegraphics[width=0.45\textwidth]{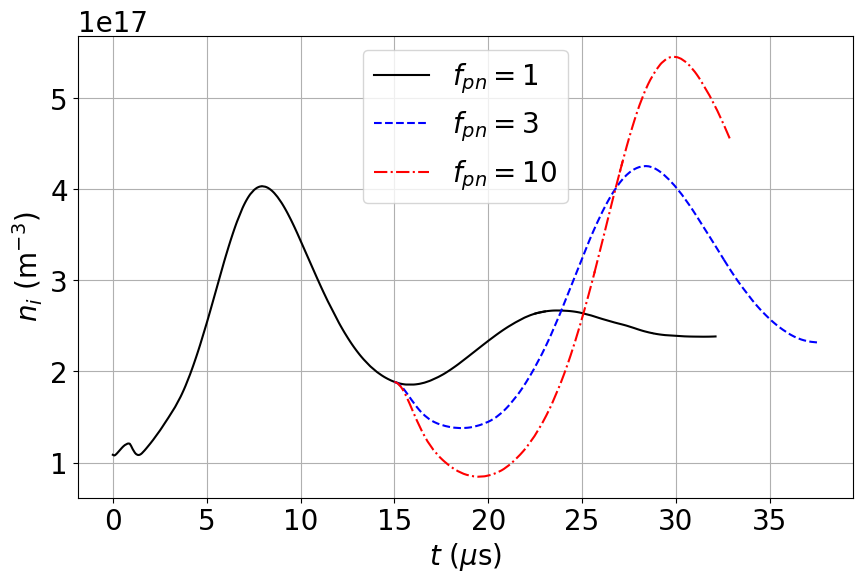}
\caption{
Results of the mean ion number density
over time of three cases with
different $f_{pn}$
of the Strong-B case.
}
\label{fig:ni}
\end{figure}

\begin{figure}[ht]
\centering
\includegraphics[width=0.45\textwidth]{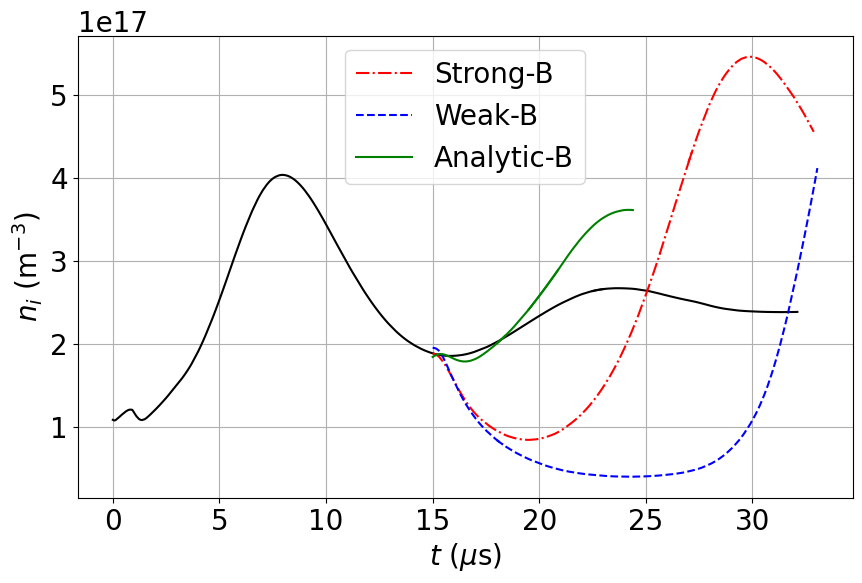}
\caption{
Results of the mean ion number density
over time of three cases with
$f_{pn}=10$ but
different B fields.
The black solid curve is a reference
of the Strong-B case with $f_{pn}=1$.
}
\label{fig:ni2}
\end{figure}

The spatial discretization uses uniform cell sizes of $\Delta x = \Delta y = \Delta z = 0.1$ mm, with a time step of $\Delta t = 1.5 \times 10^{-11}$ s. The channel region employs $N_{xc} = 128$ and $N_{zc} = 72$ cells, corresponding to physical dimensions of $L_{xc} = 1.28$ cm and $L_{zc} = 0.72$ cm. The plume region utilizes $N_{xp} = 256$ and $N_{zp} = 184$ cells, with physical extents of $L_{xp} = 2.56$ cm and $L_{zp} = 1.84$ cm. Both regions share identical azimuthal dimensions with $N_y = 64$ and $L_{yc} = L_{yp} = L_y = 0.64$ cm.
While the previous study \cite{zhong20253dpicstudymagnetic} employed a macro-particle weight of $w_0 \approx 19555$, the current work reduces this weight by factors of $f_{pn} = 3$ and $10$ to simulate three and ten times more macro-particles, respectively, thereby mitigating numerical noise as done in \cite{IEPC-2025-063}.

Note that the chosen cell size in this work
is twice as that used in the two benchmarks
\cite{Charoy2019,Villafana2021},
and could be over the local Debye length
during the simulation when the electron number density $n_e$ becomes high
and the electron temperature $T_e$ becomes low.
The timestep is set to be $1.5 \times 10^{-11}$ s
as the same as that in \cite{Villafana2021},
but bigger than $5 \times 10^{-12}$ s applied in \cite{Charoy2019,Villafana2023}.
However, the chosen cell size and timestep
are the best we can do so far for carrying out such computationally
expensive 3D PIC simulations
within limit resources.
Because of that, detailed spectral analyses are not done,
which would require higher accuracy of the simulation,
instead macroscopic quantities averaged in space and time are discussed more
in this work.
In addition, note that the domain length in the azimuthal direction is
set to be 0.64 cm with 64 cells to reduce the computational cost,
which should be large enough to cover the major waves we are focusing on,
and its effects are found to be negligible compared to 1.28 cm as
discussed in a previous work \cite{Xie_2024}.

There were three simulation cases labeled
Analytic-B, Strong-B, and Weak-B
studied in the previous paper
\cite{zhong20253dpicstudymagnetic}.
Analytic-B applied the commonly used analytic
B field with only the radial component
\cite{Villafana2021,Villafana2023,Xie_2024};
Strong-B and Weak-B applied a more realistic
magnetic field configuration generated by 
the FEMM software (www.femm.info);
Strong-B has an about 200 Gauss peak value on the centerline
along $z$,
while Weak-B has about 100 Gauss,
as shown in Fig.\ref{fig:B}.
In this paper, we continue to study these three
cases, and load their results at
$10^6$ time steps (15 $\mu$s)
as initialization with a scaling factor
$f_{pn}$.
If $f_{pn}>1$,
each macro-particle loaded will be copied
$f_{pn}$ times with the same position
and velocity, such that there are
more macro-particles at the initialization,
and the particle weight will be reduced
by $f_{pn}$ accordingly.
The simulation results of the ion number density
over time of the Strong-B case are shown in Fig.\ref{fig:ni}.
As can be seen, increasing the number of macro-particles
modifies the $n_i$ evolution to some extent,
and we chose the most macro-particles that we can apply so far with $f_{pn}=10$
for running the other two cases in order to obtain less noisy results.
The $n_i$ evolution of the three cases with $f_{pn}=10$ are shown
in Fig.\ref{fig:ni2},
from which different stages of the ionization breathing-mode can be identified
and will be referred to later in the paper.
Other parts of the simulation methods,
such as the cathode injection,
the MCC ionization, and
the fluid solver on the continuity equation
are the same as
the previous paper 
\cite{zhong20253dpicstudymagnetic,IEPC-2025-063}.

\section{Simulation Results}
\label{sec:results}

Because the simulations considered in this paper
were also studied in a previous conference paper
\cite{IEPC-2025-063},
in which the commonly presented
instantaneous fields of $n_e$, $n_i$, and $E_y$
were discussed already,
these fields are not shown in this paper
redundantly.
In addition, analyses were also presented
for the three corresponding simulation cases
with less number of macro-particles
($f_{pn}=1$) in another previous paper
\cite{zhong20253dpicstudymagnetic},
therefore this paper only focuses
on the discussion of fields
directly related to the electron transport
due to the instability
under different magnetic fields.

\begin{figure*}[ht]
\centering
\includegraphics[width=0.3\textwidth]{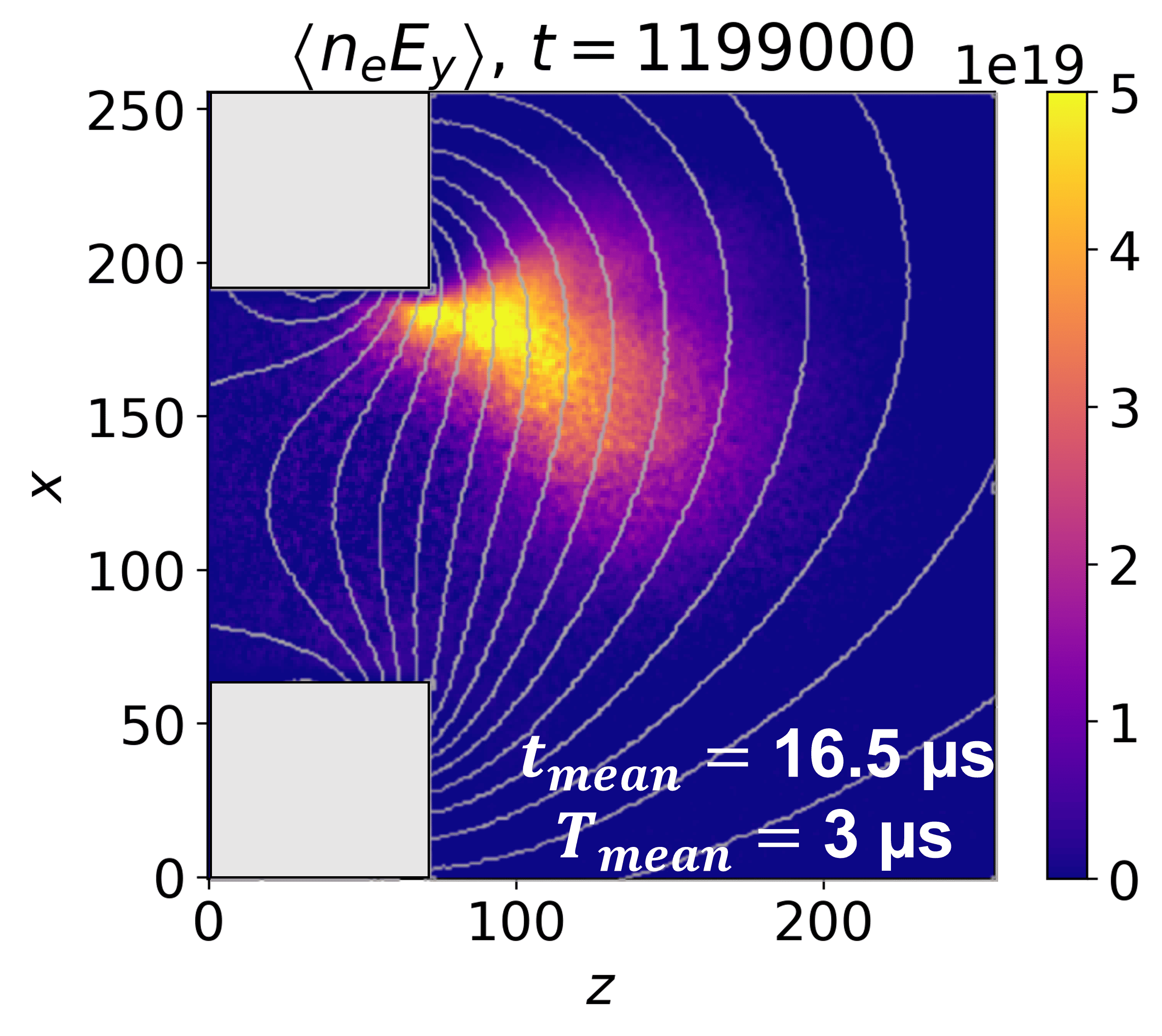}
\includegraphics[width=0.3\textwidth]{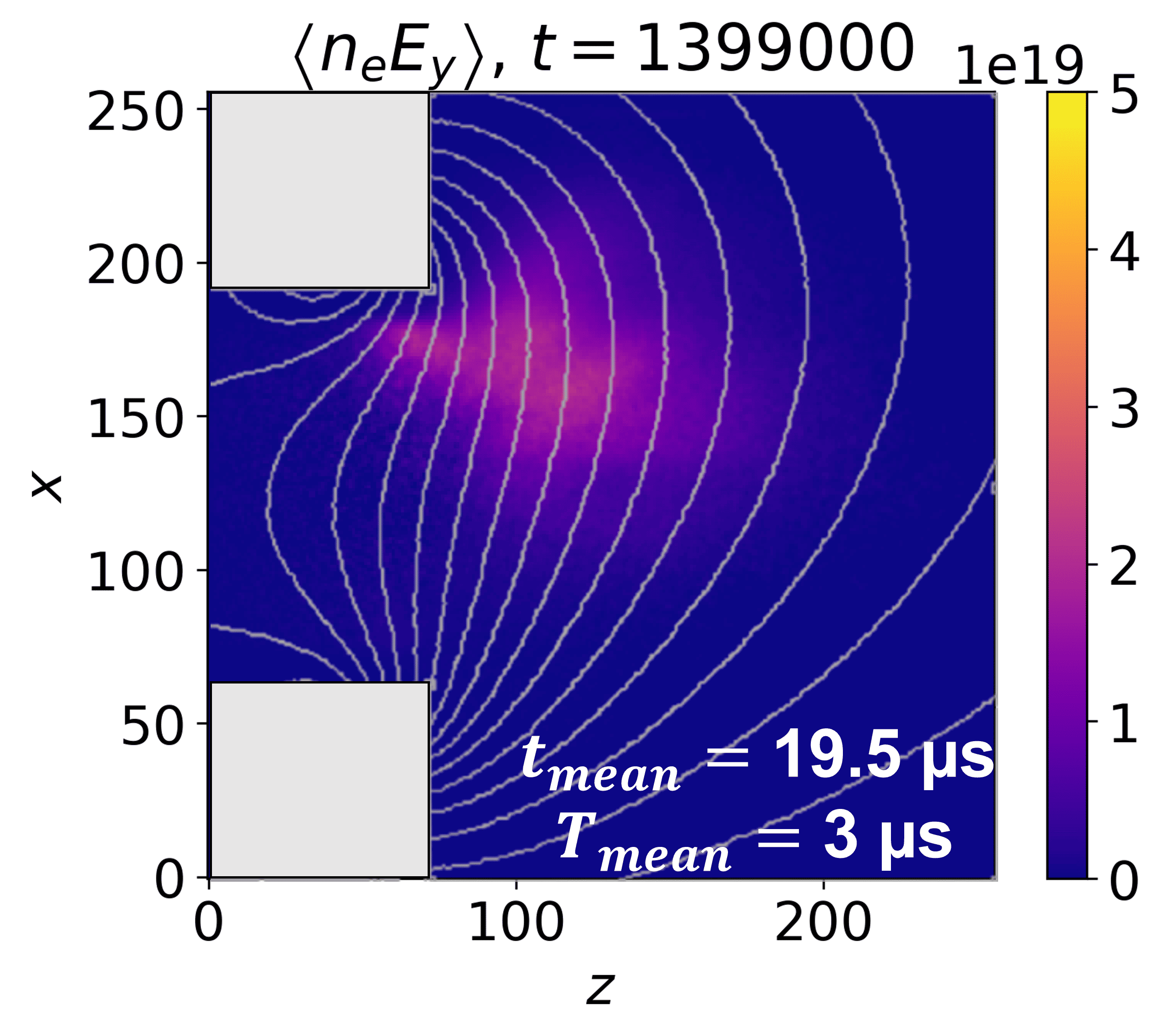}
\includegraphics[width=0.3\textwidth]{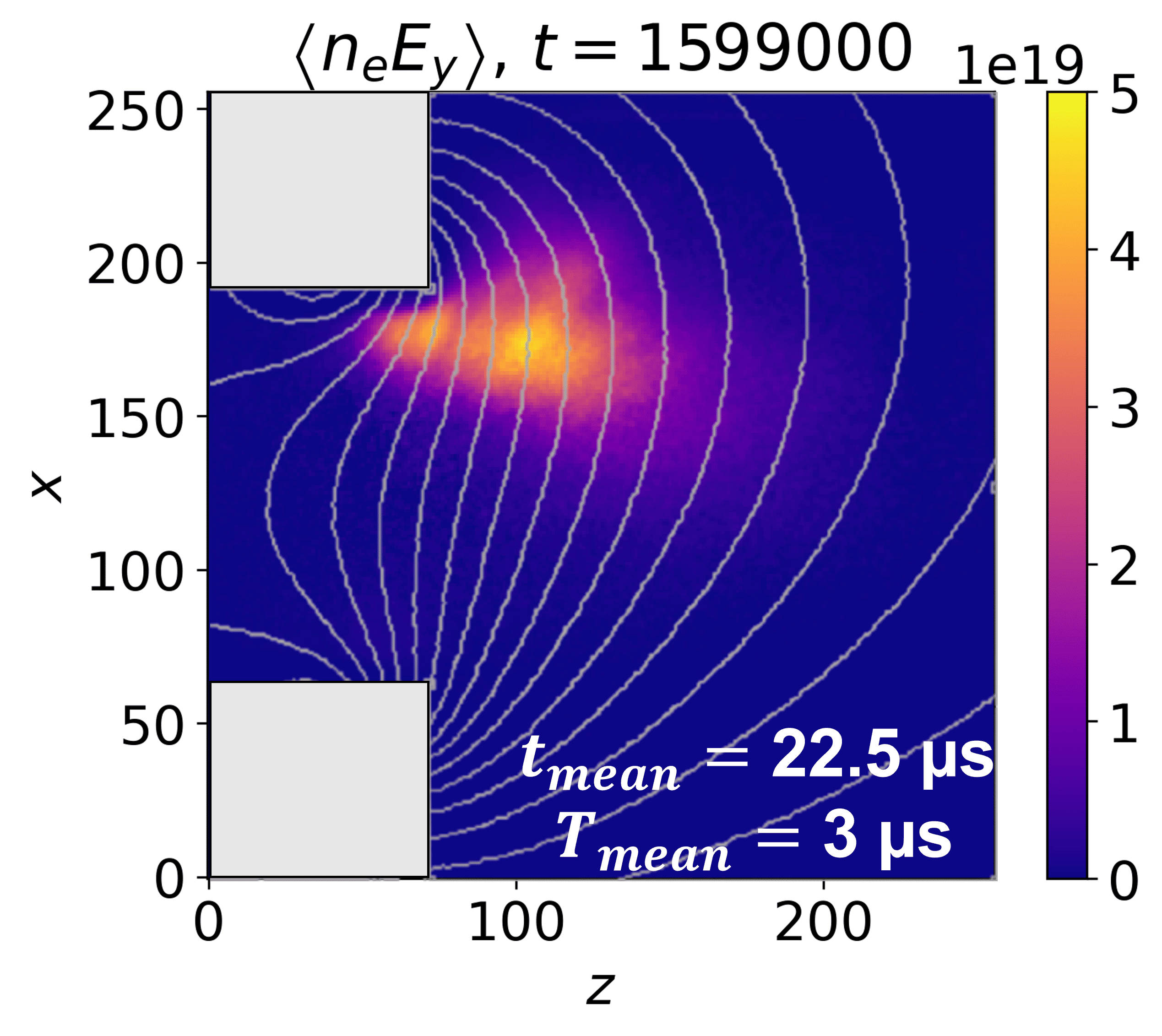}\\
\includegraphics[width=0.3\textwidth]{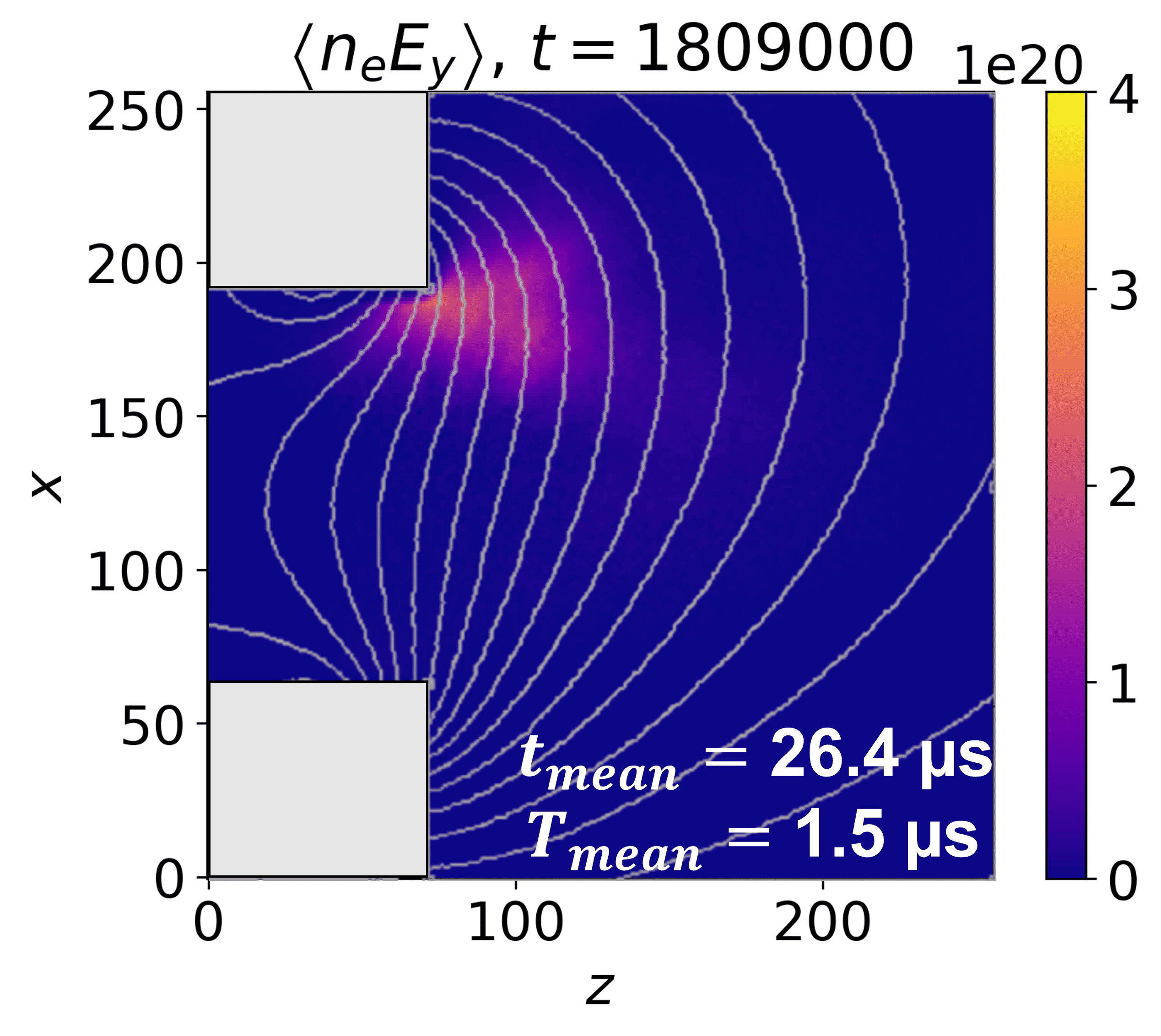}
\includegraphics[width=0.3\textwidth]{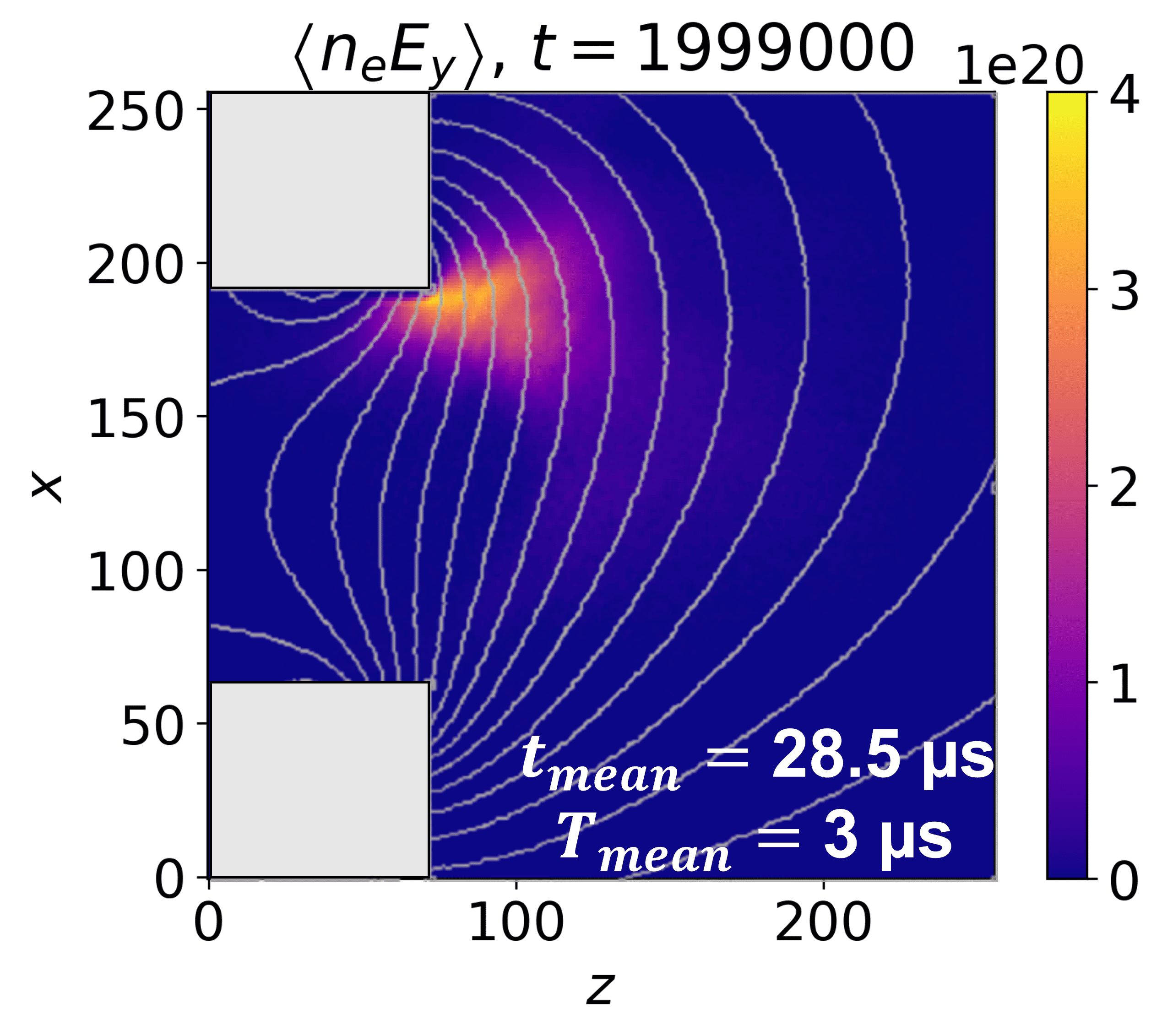}
\includegraphics[width=0.3\textwidth]{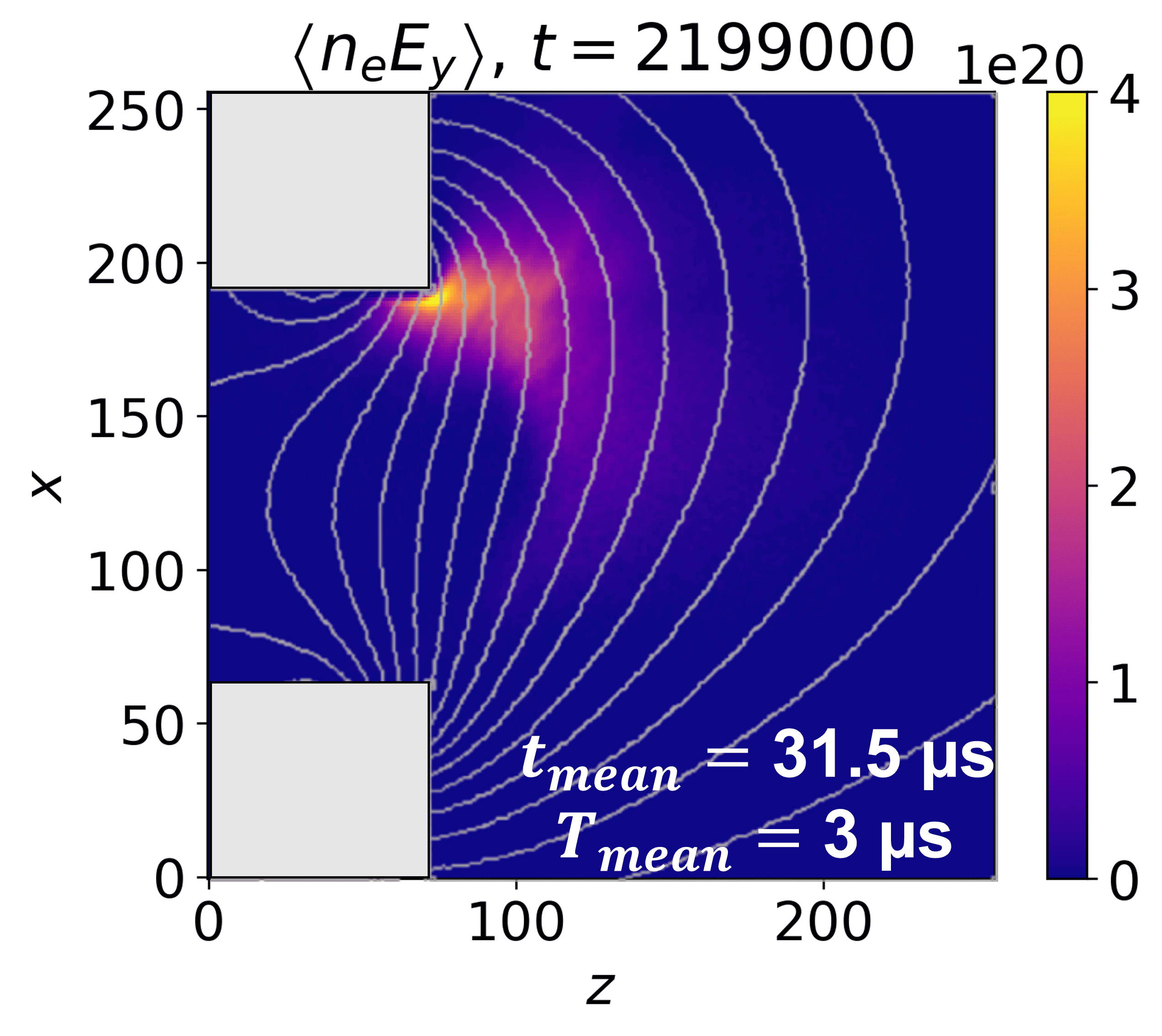}
\caption{
The correlation between the electron density and the azimuthal
electric field of the Strong-B case
evaluated using different time periods.
}
\label{fig:nE_01}
\end{figure*}

\begin{figure*}[ht]
\centering
\includegraphics[width=0.3\textwidth]{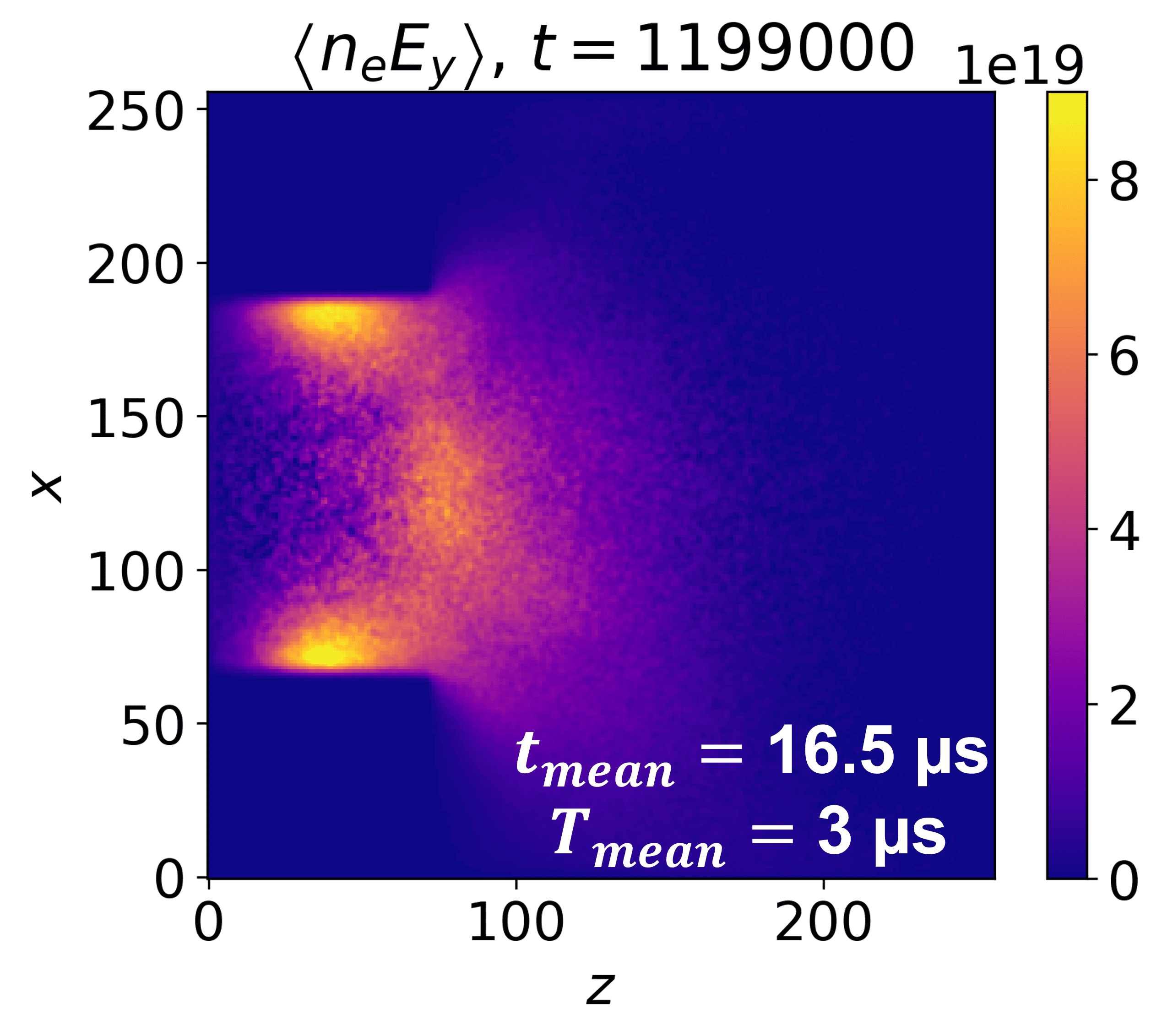}
\includegraphics[width=0.3\textwidth]{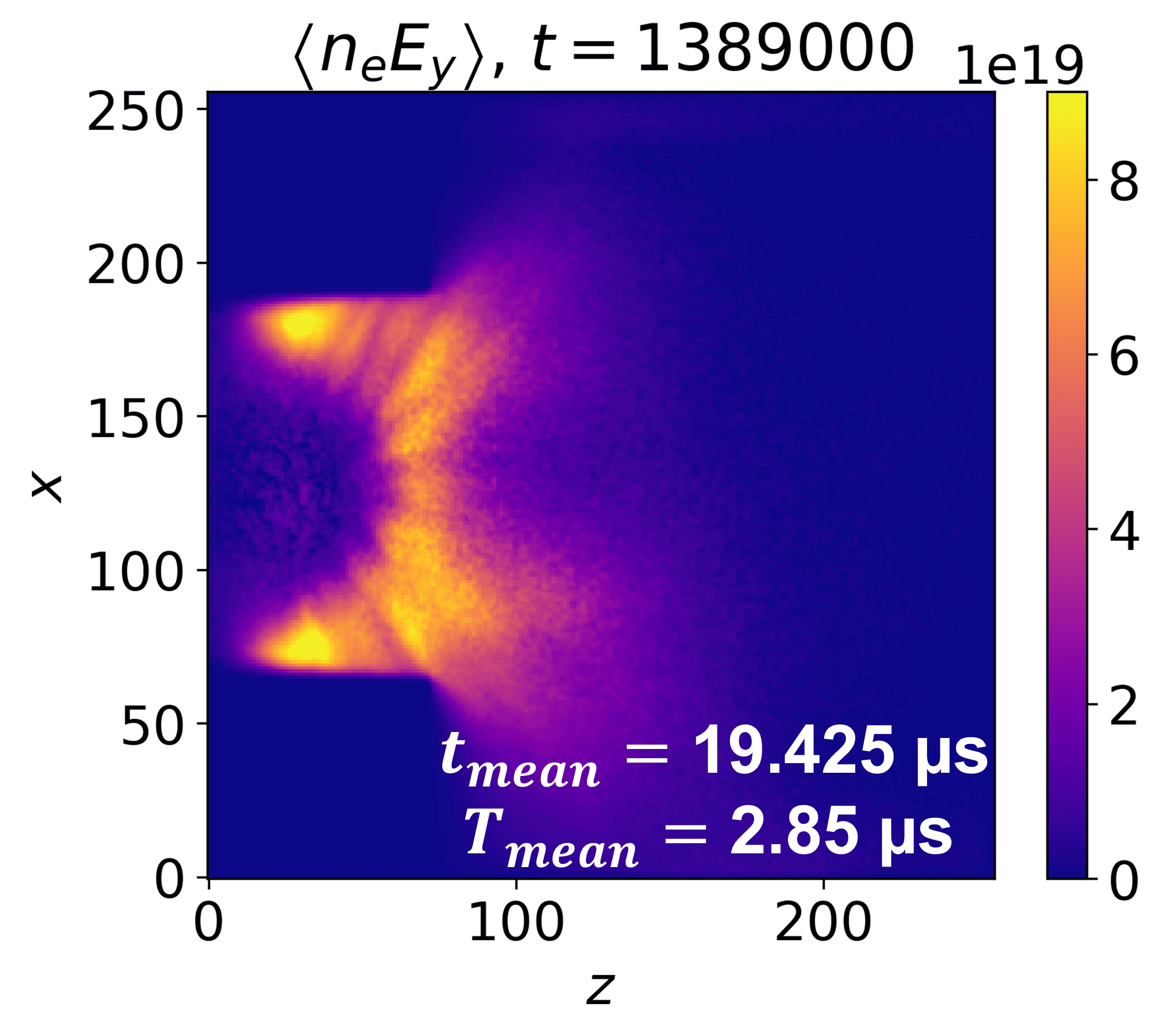}
\includegraphics[width=0.3\textwidth]{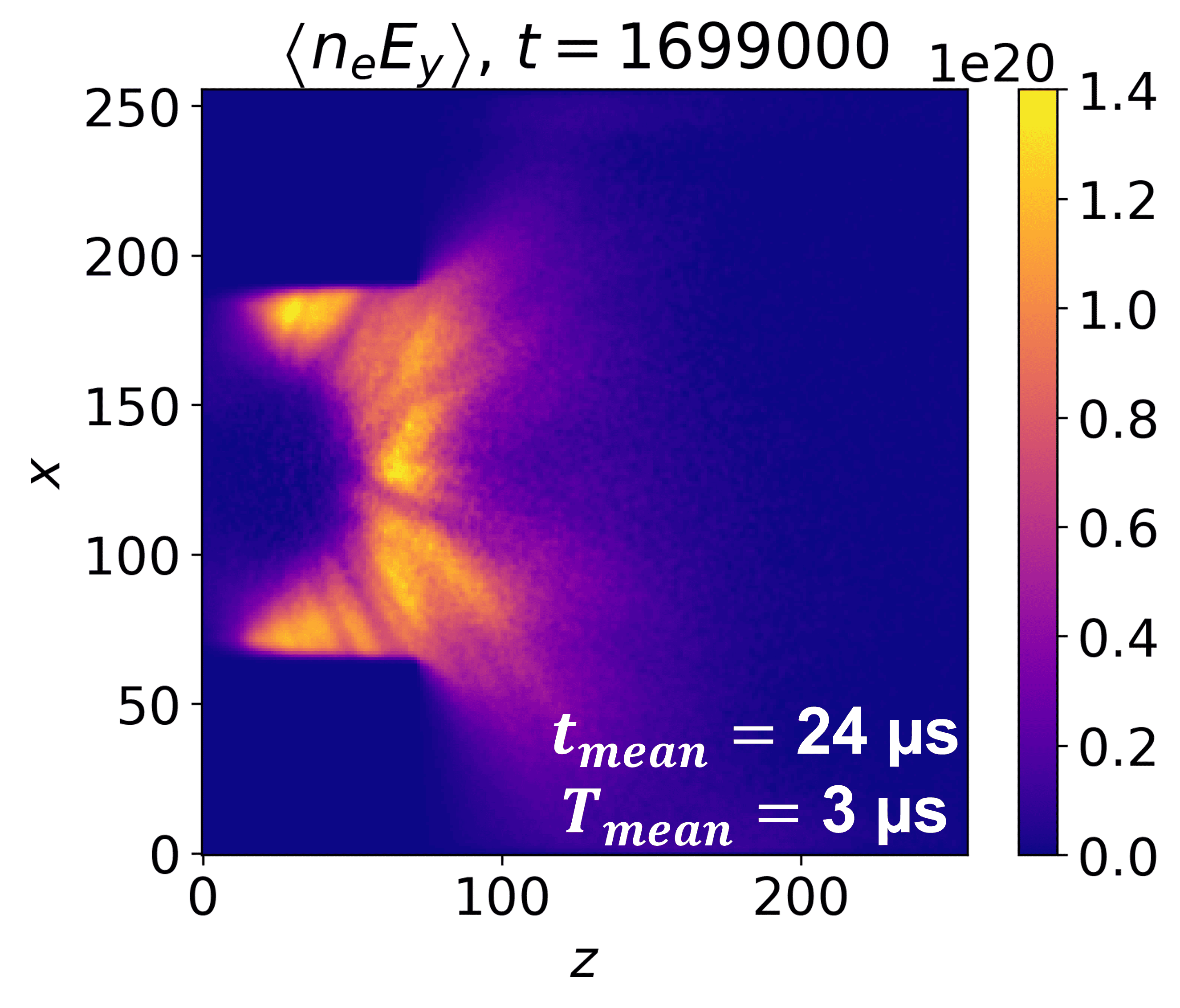}
\caption{
The correlation between the electron density and the azimuthal
electric field of the Analytic-B case
evaluated using different time periods.
}
\label{fig:nE_03}
\end{figure*}

\begin{figure*}[ht]
\centering
\includegraphics[width=0.3\textwidth]{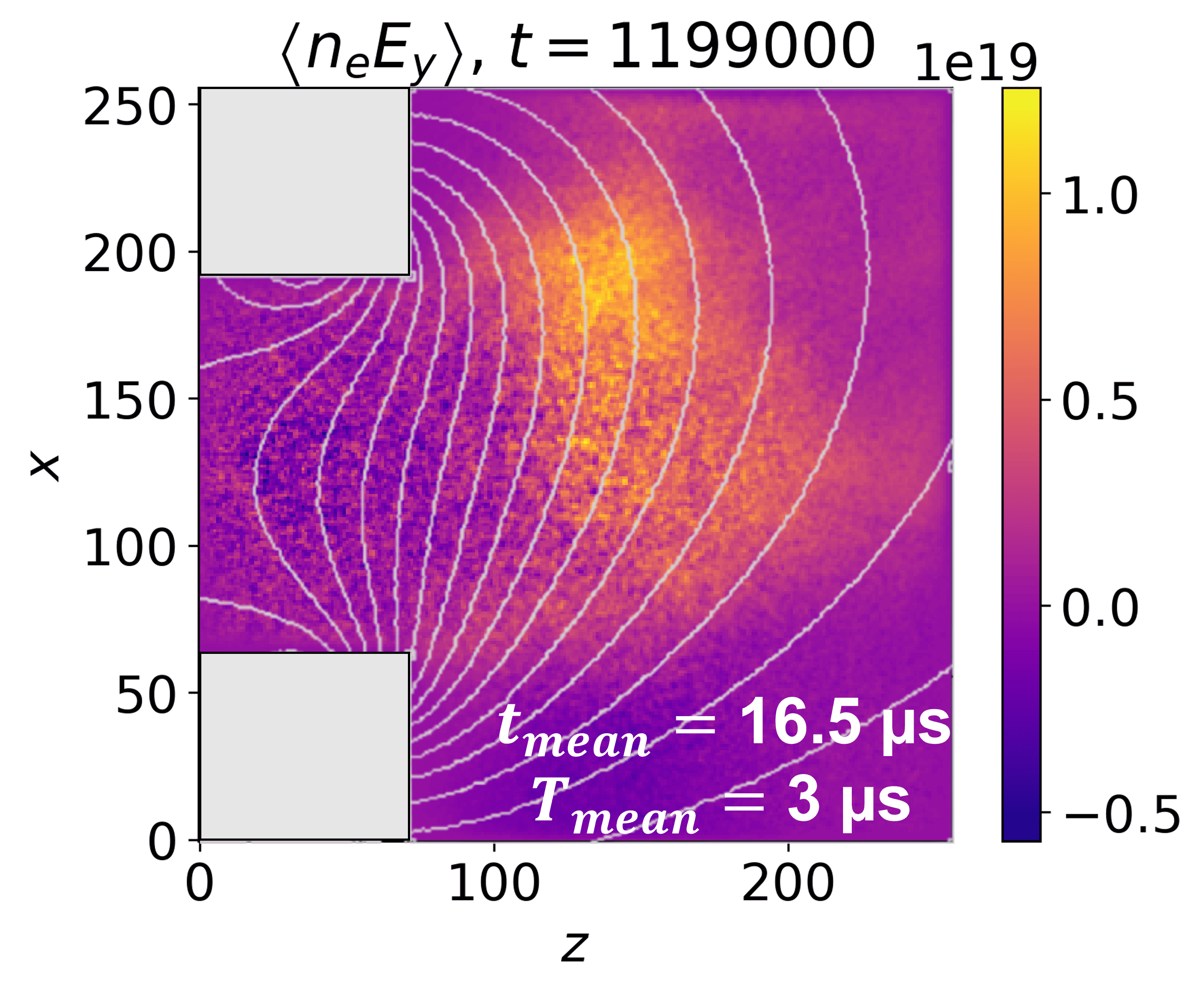}
\includegraphics[width=0.3\textwidth]{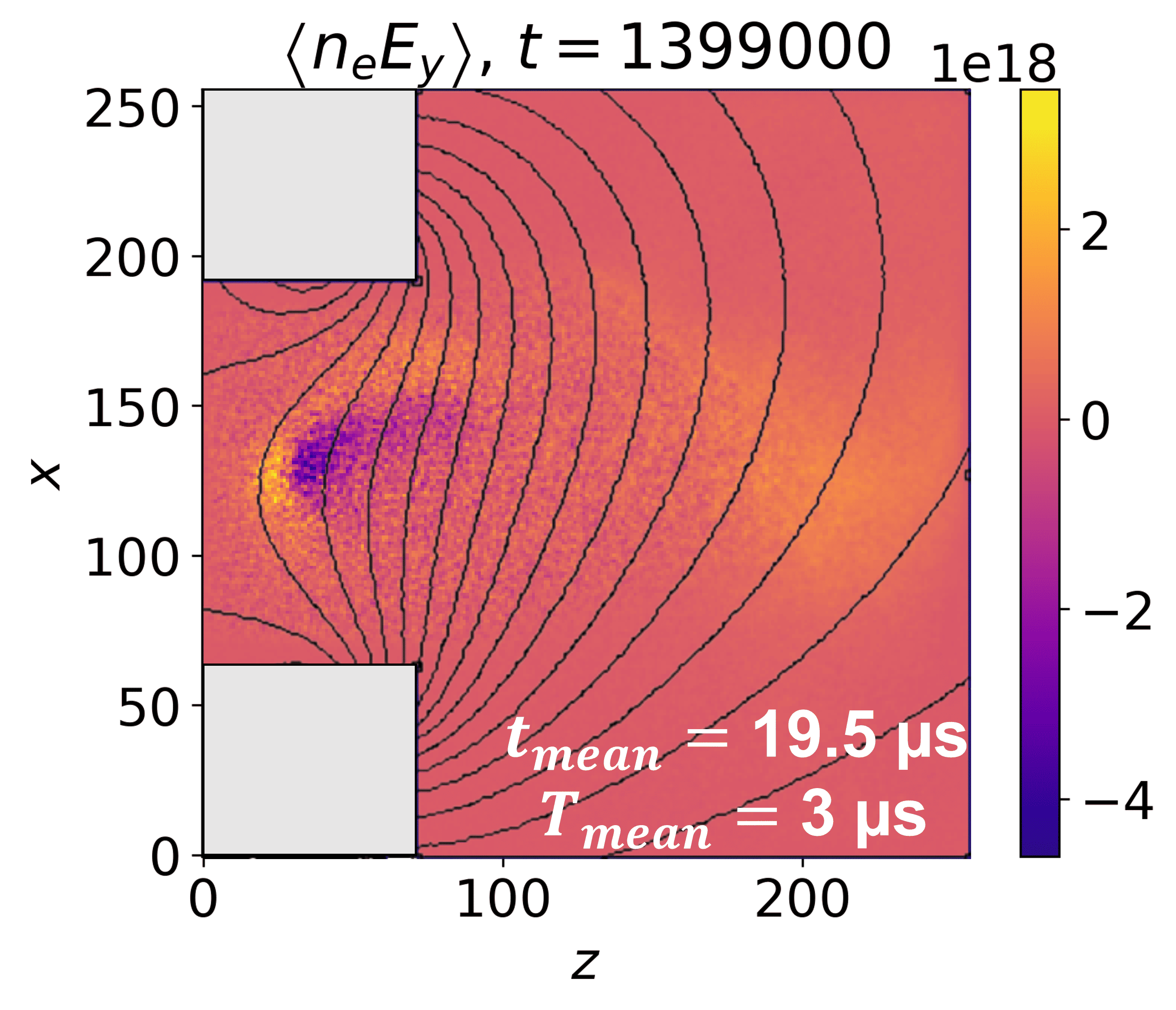}
\includegraphics[width=0.3\textwidth]{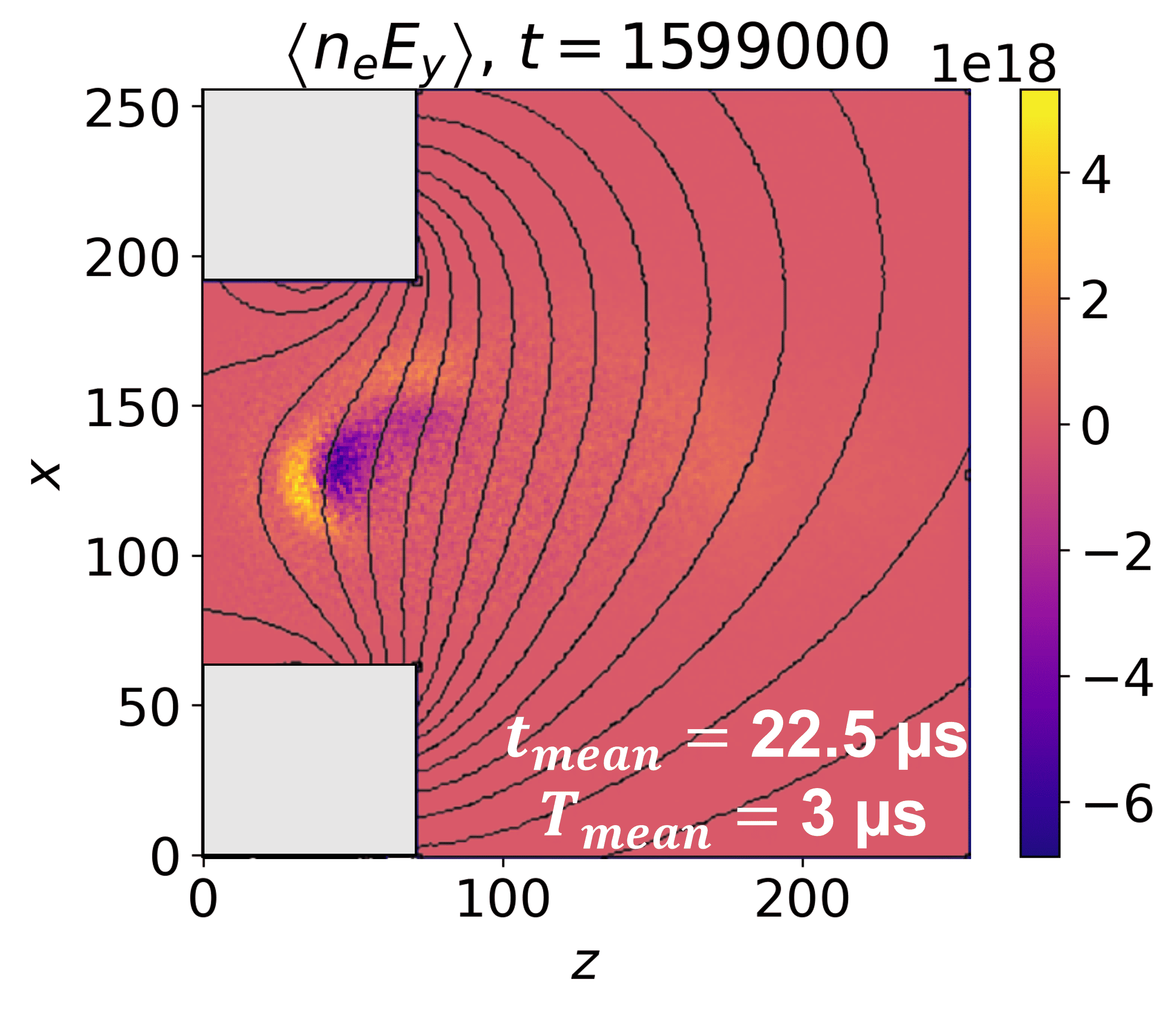}
\includegraphics[width=0.3\textwidth]{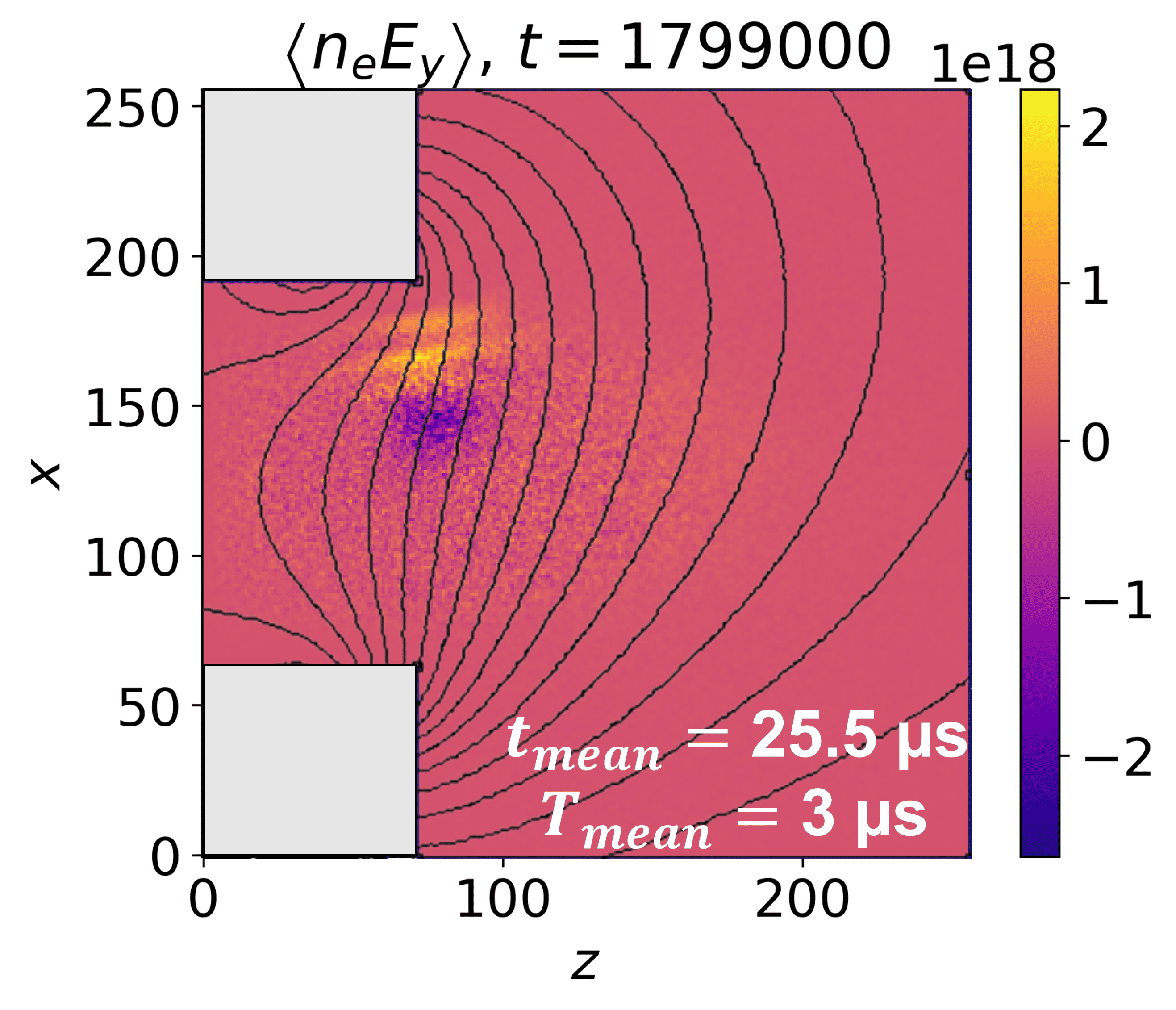}
\includegraphics[width=0.3\textwidth]{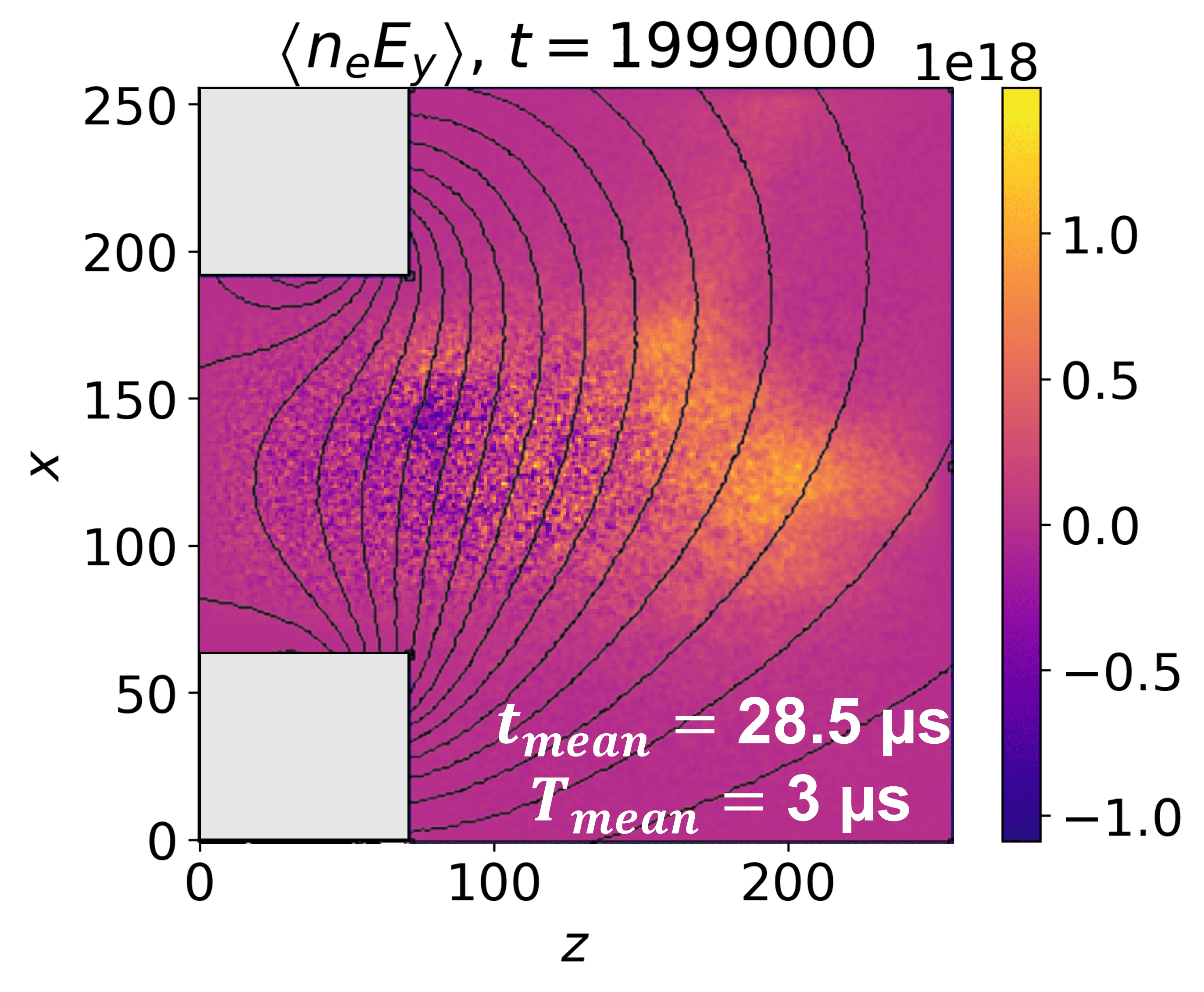}
\includegraphics[width=0.3\textwidth]{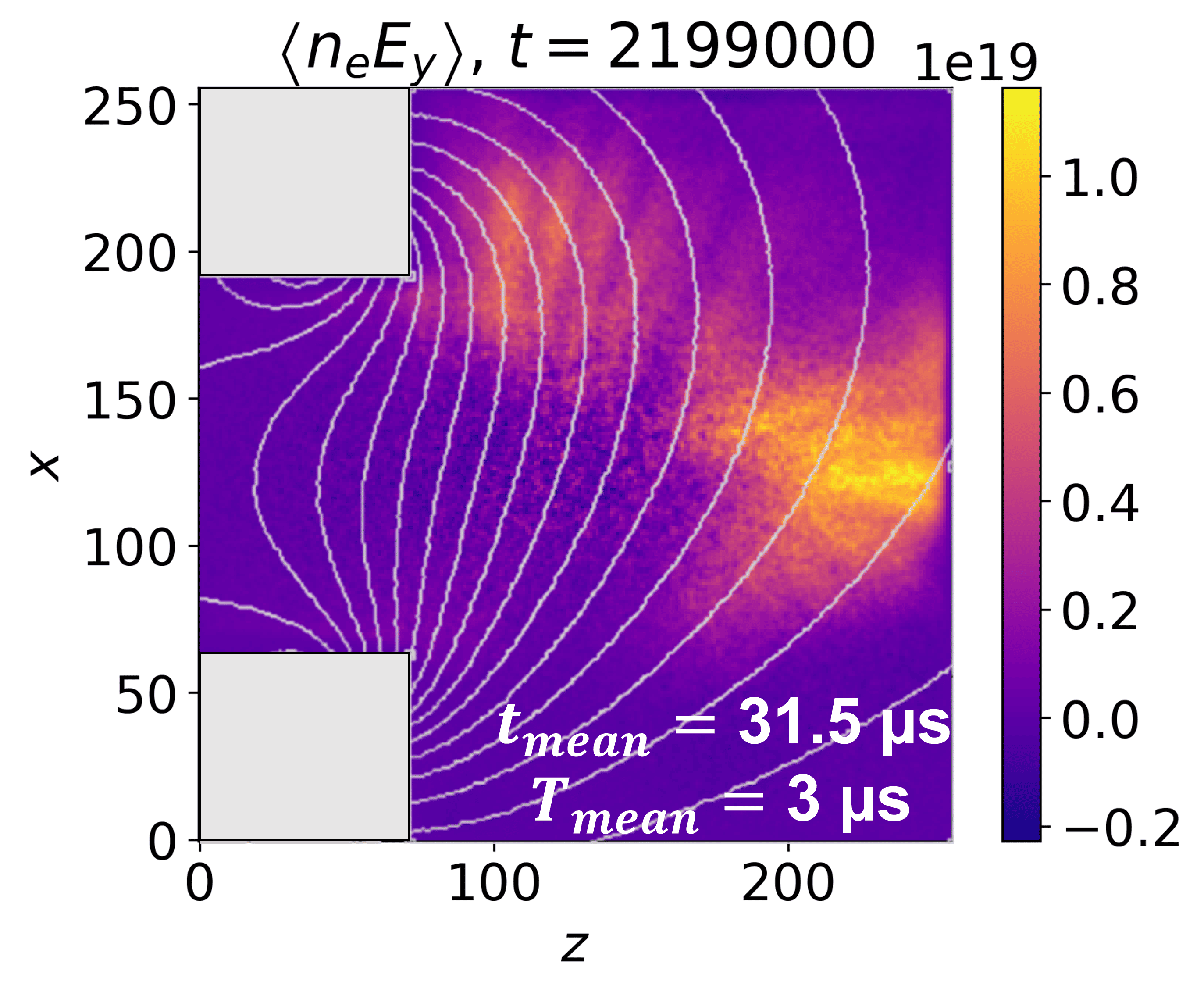}
\caption{
The correlation between the electron density and the azimuthal
electric field of the Weak-B case
evaluated using different time periods.
}
\label{fig:nE_04}
\end{figure*}

\begin{figure}[ht]
\centering
\includegraphics[width=0.3\textwidth]{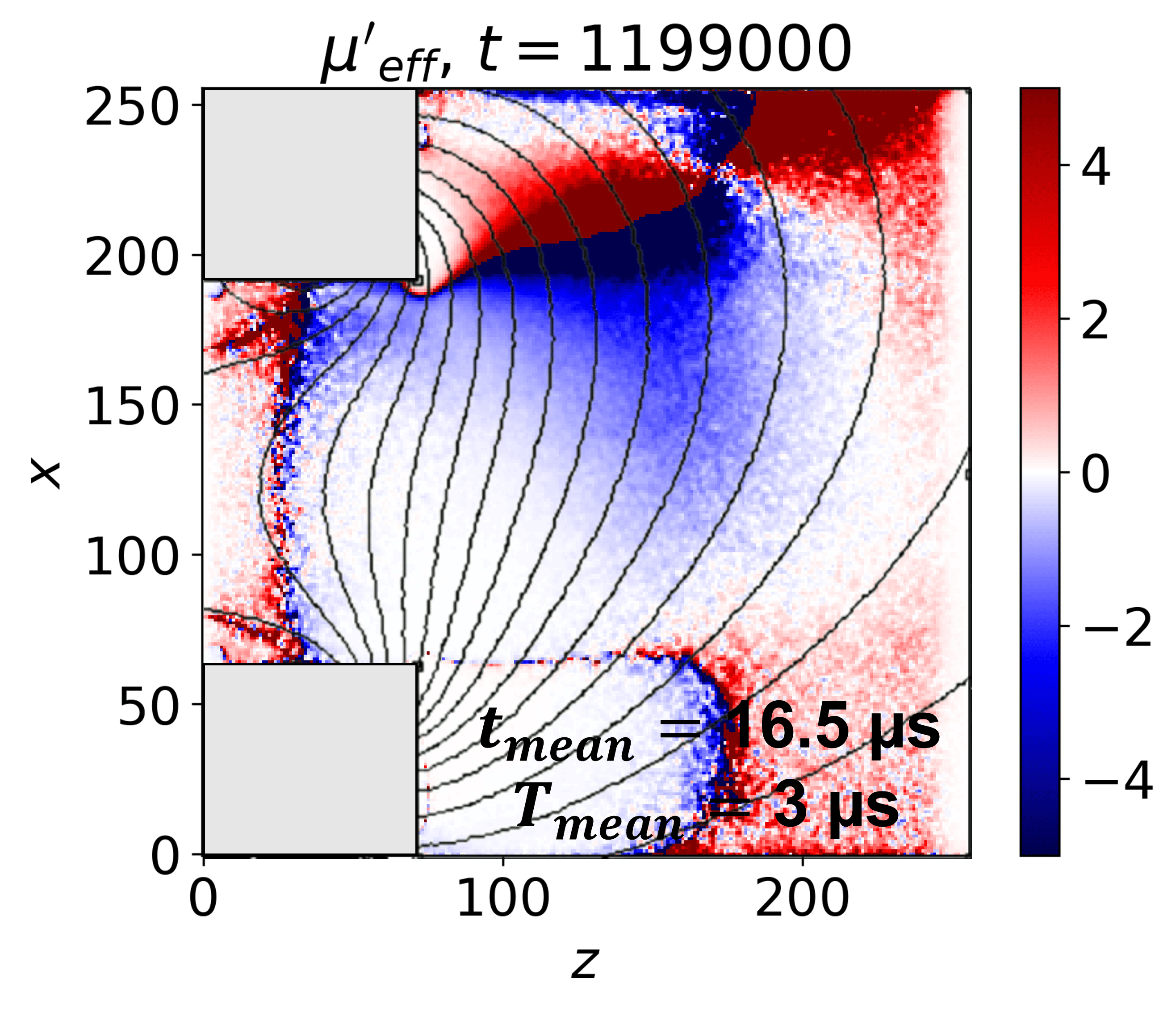}
\includegraphics[width=0.3\textwidth]{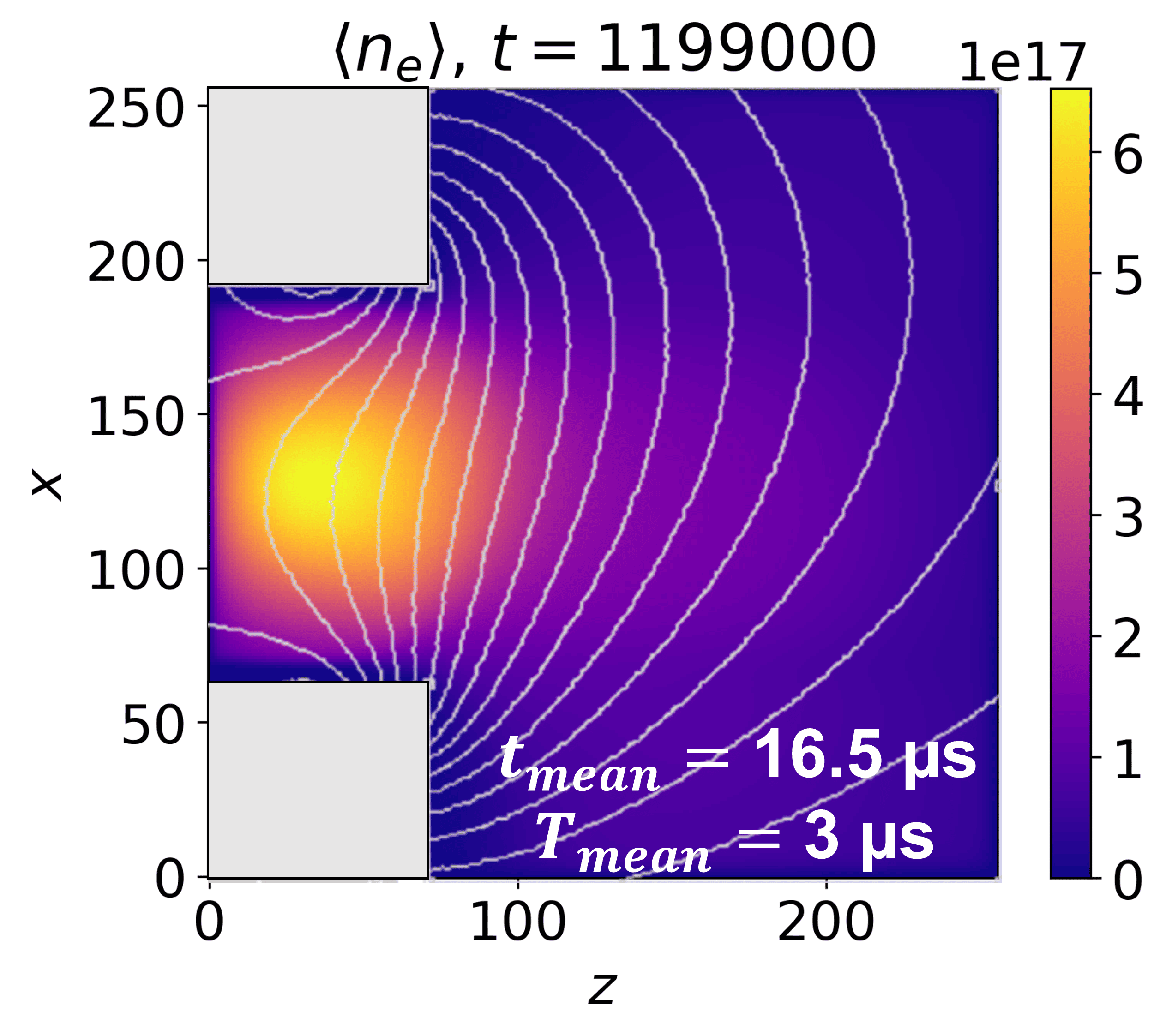}
\includegraphics[width=0.3\textwidth]{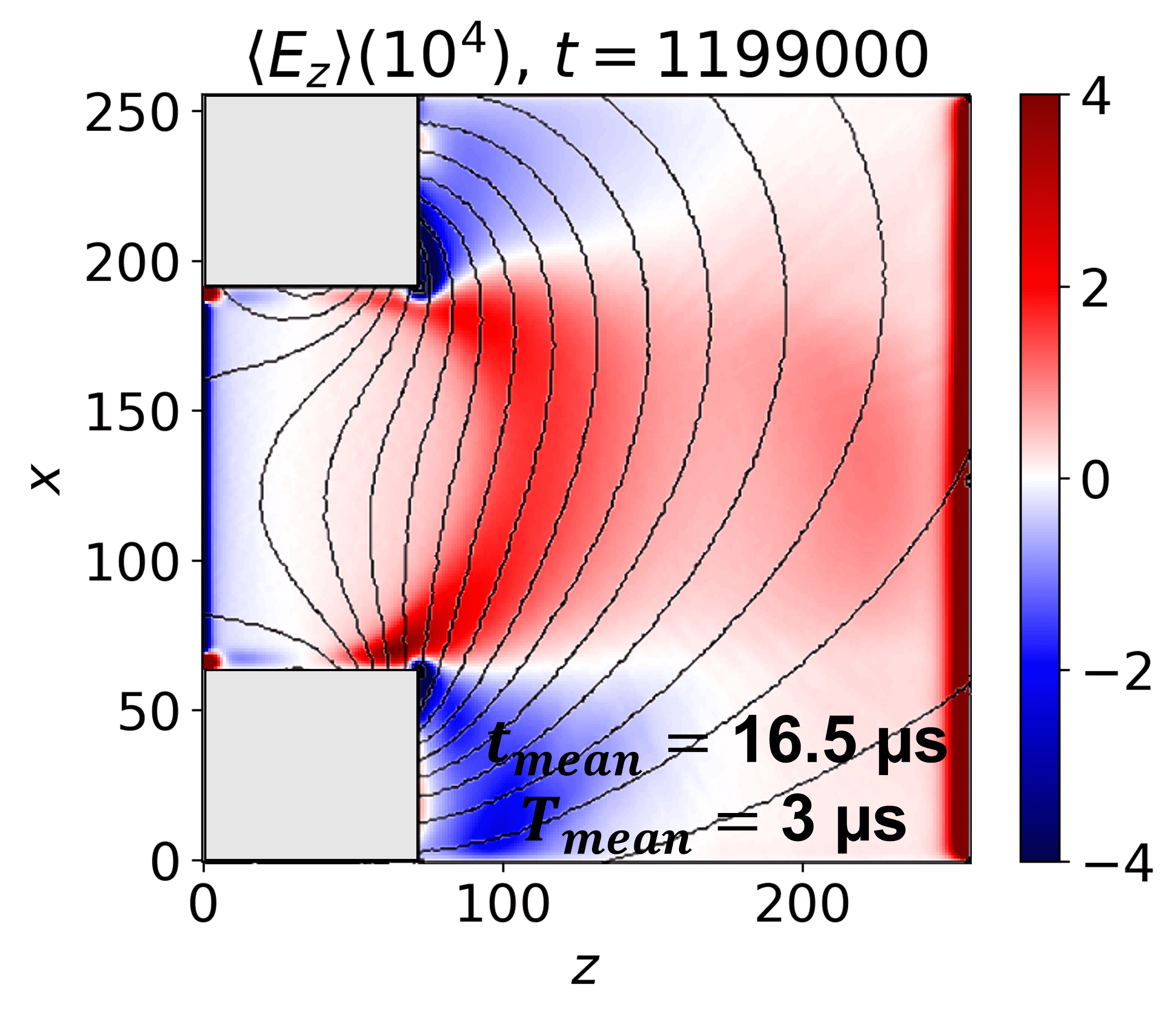}\\
\includegraphics[width=0.3\textwidth]{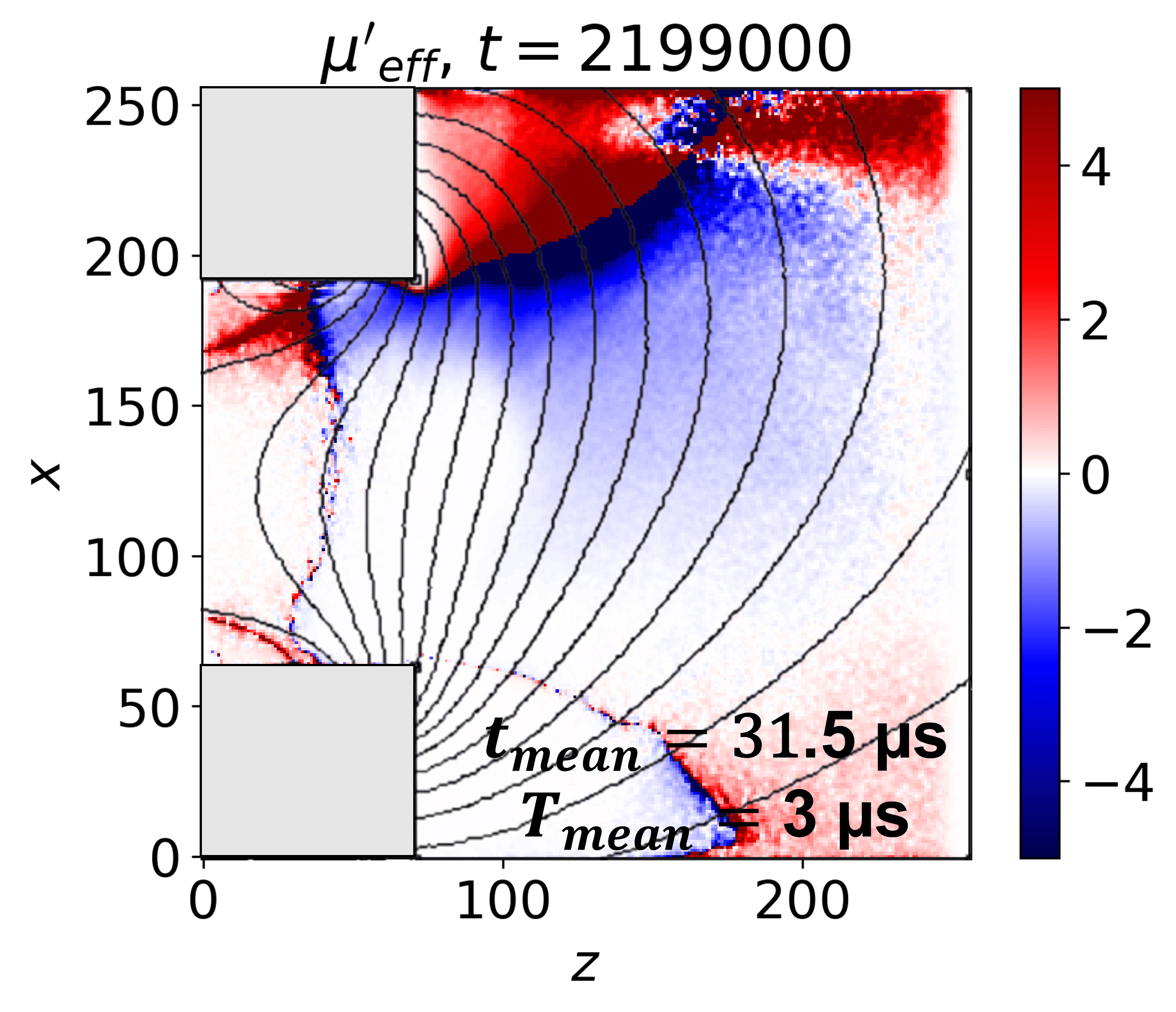}
\includegraphics[width=0.3\textwidth]{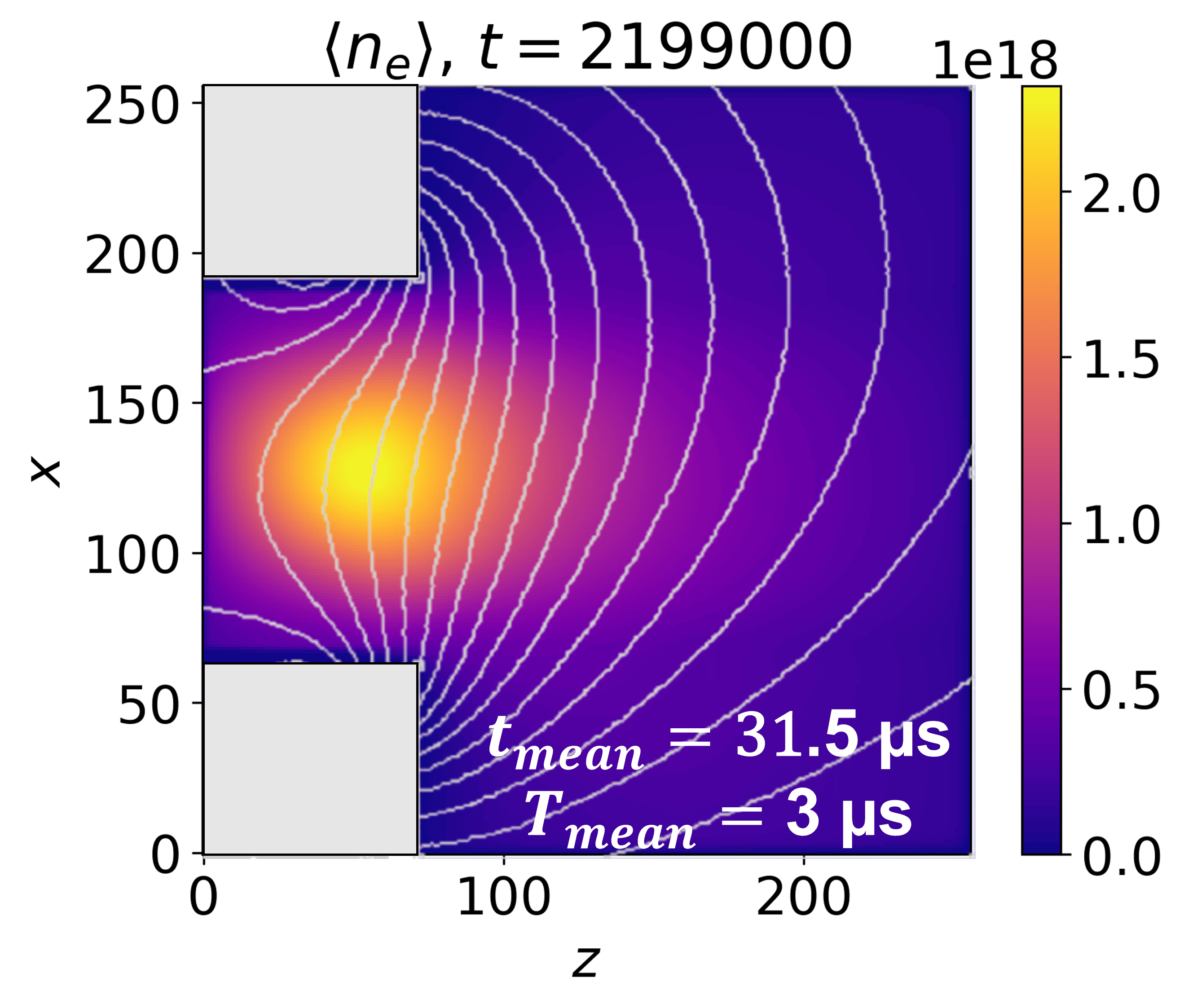}
\includegraphics[width=0.3\textwidth]{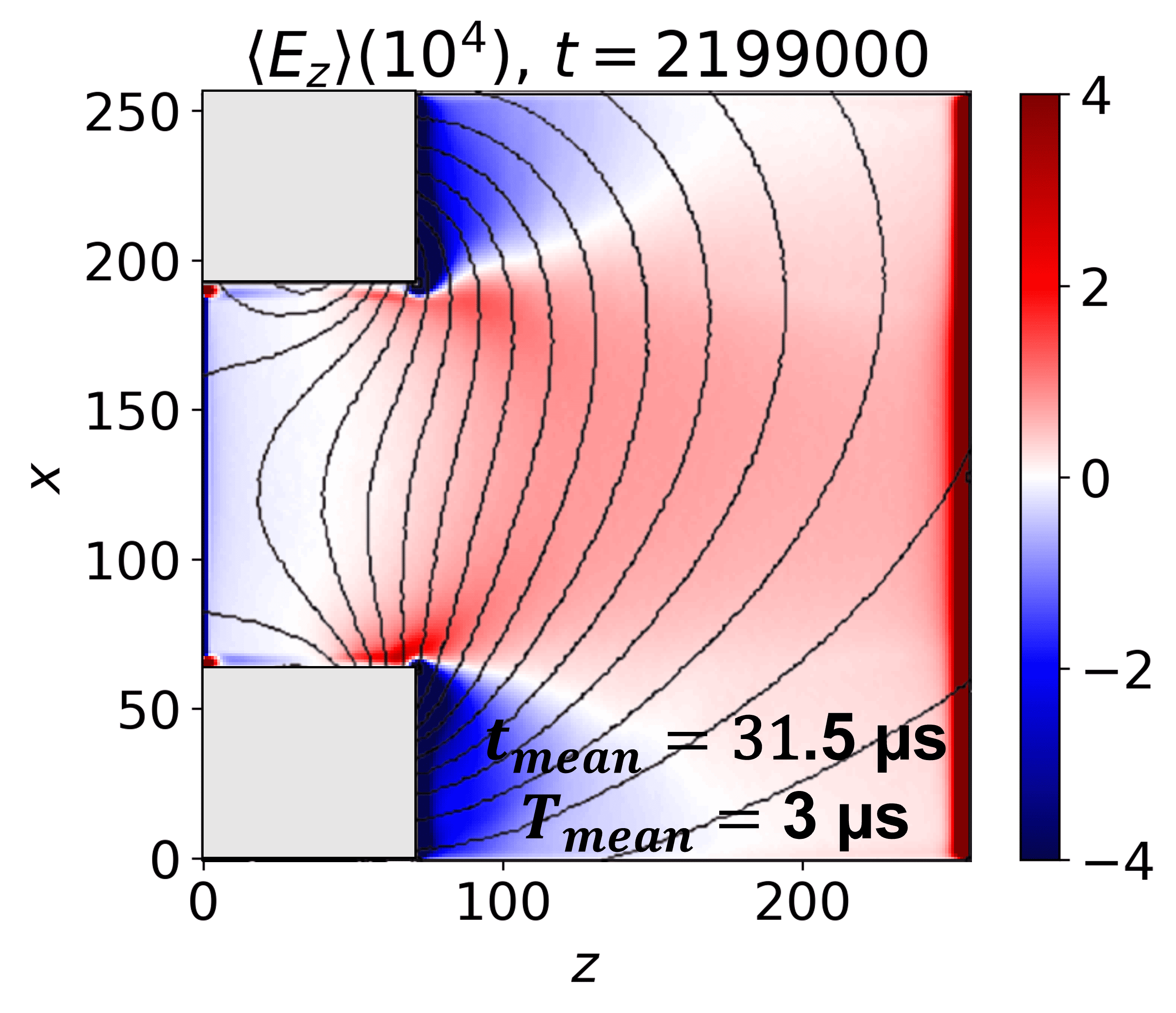}
\caption{
The effective electron mobility, averaged electron density, averaged axial electric field
of the Strong-B case
evaluated using different time periods.
}
\label{fig:mu_1199000}
\end{figure}

\begin{figure}[ht]
\centering
\includegraphics[width=0.3\textwidth]{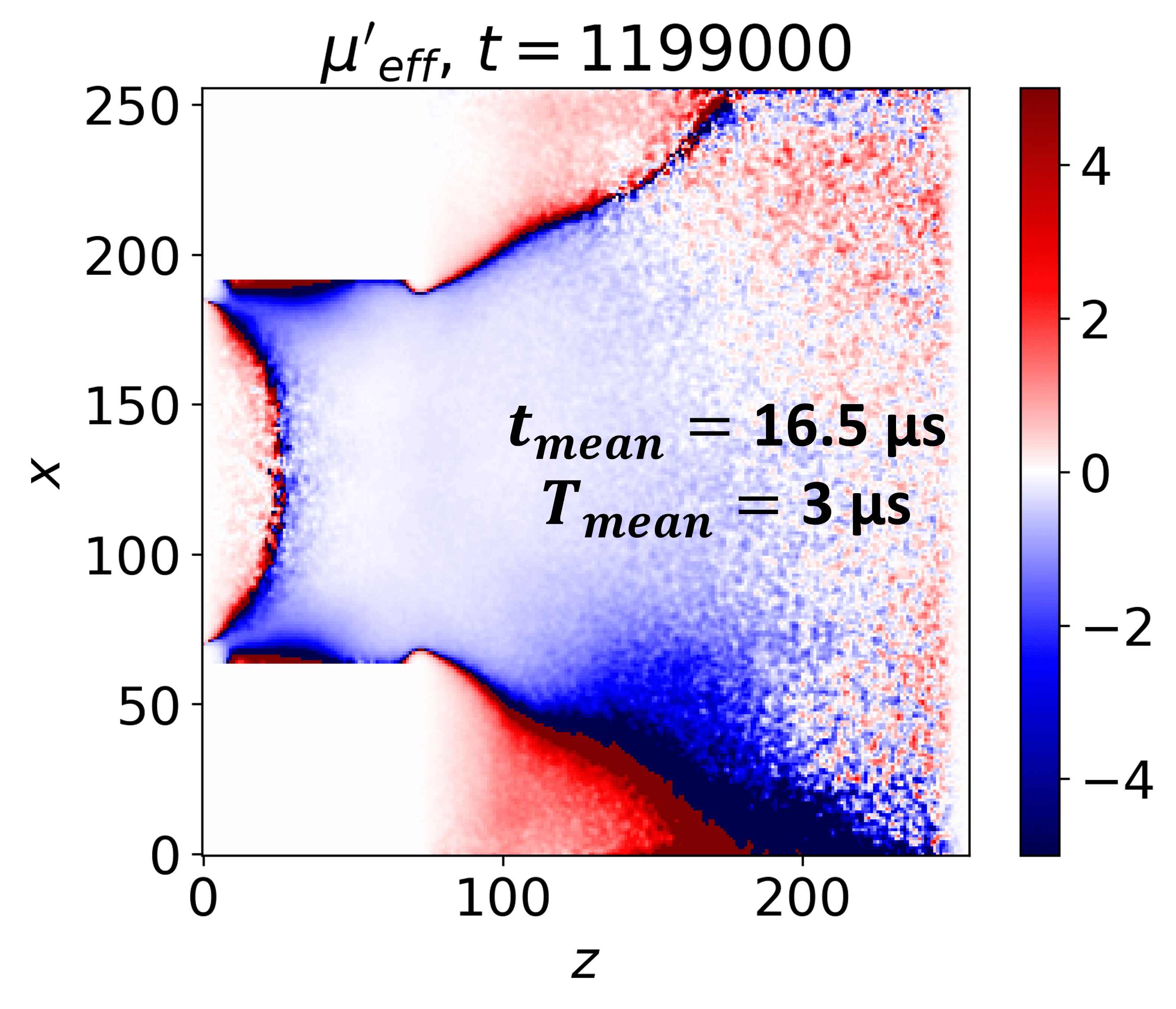}
\includegraphics[width=0.3\textwidth]{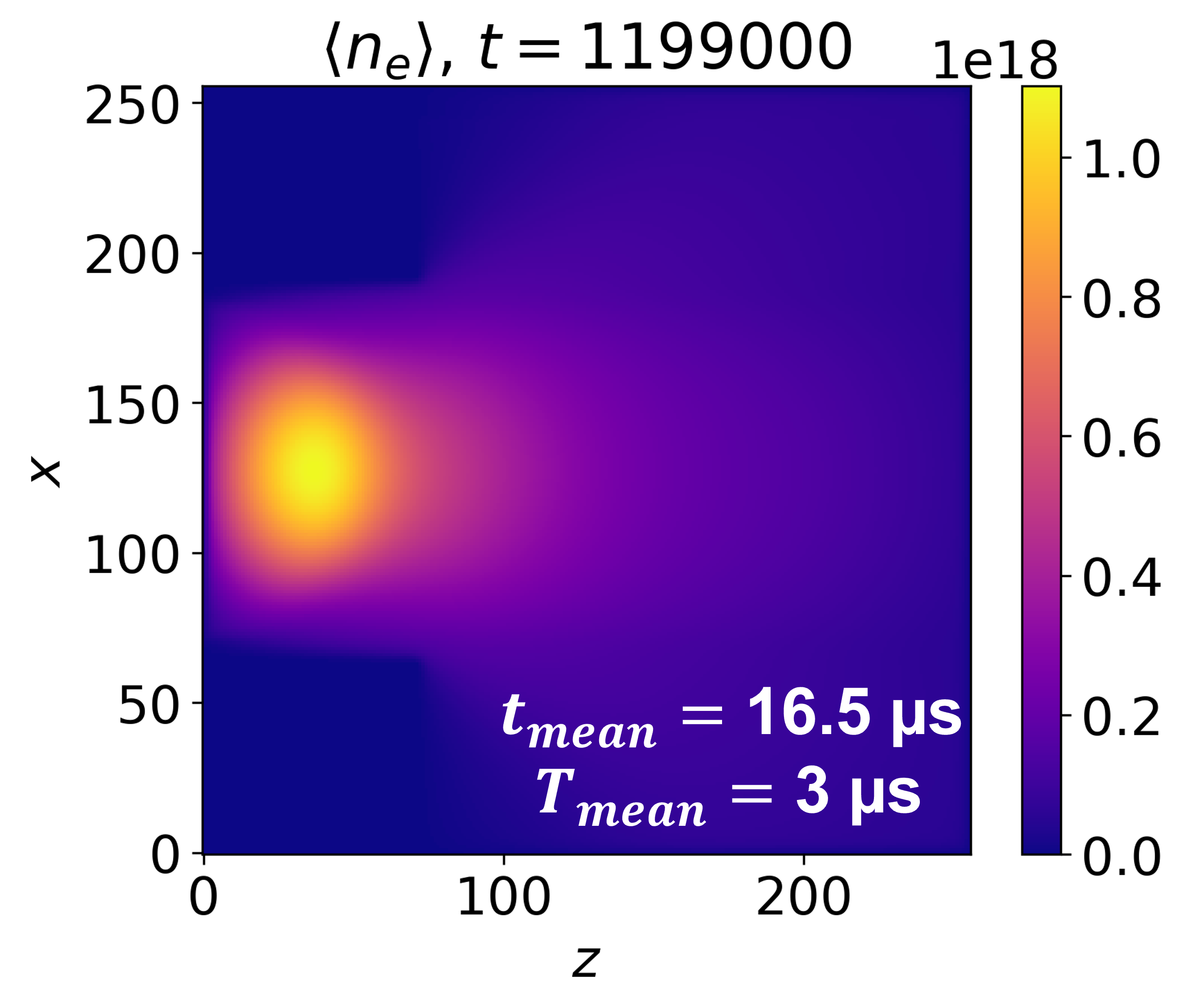}
\includegraphics[width=0.3\textwidth]{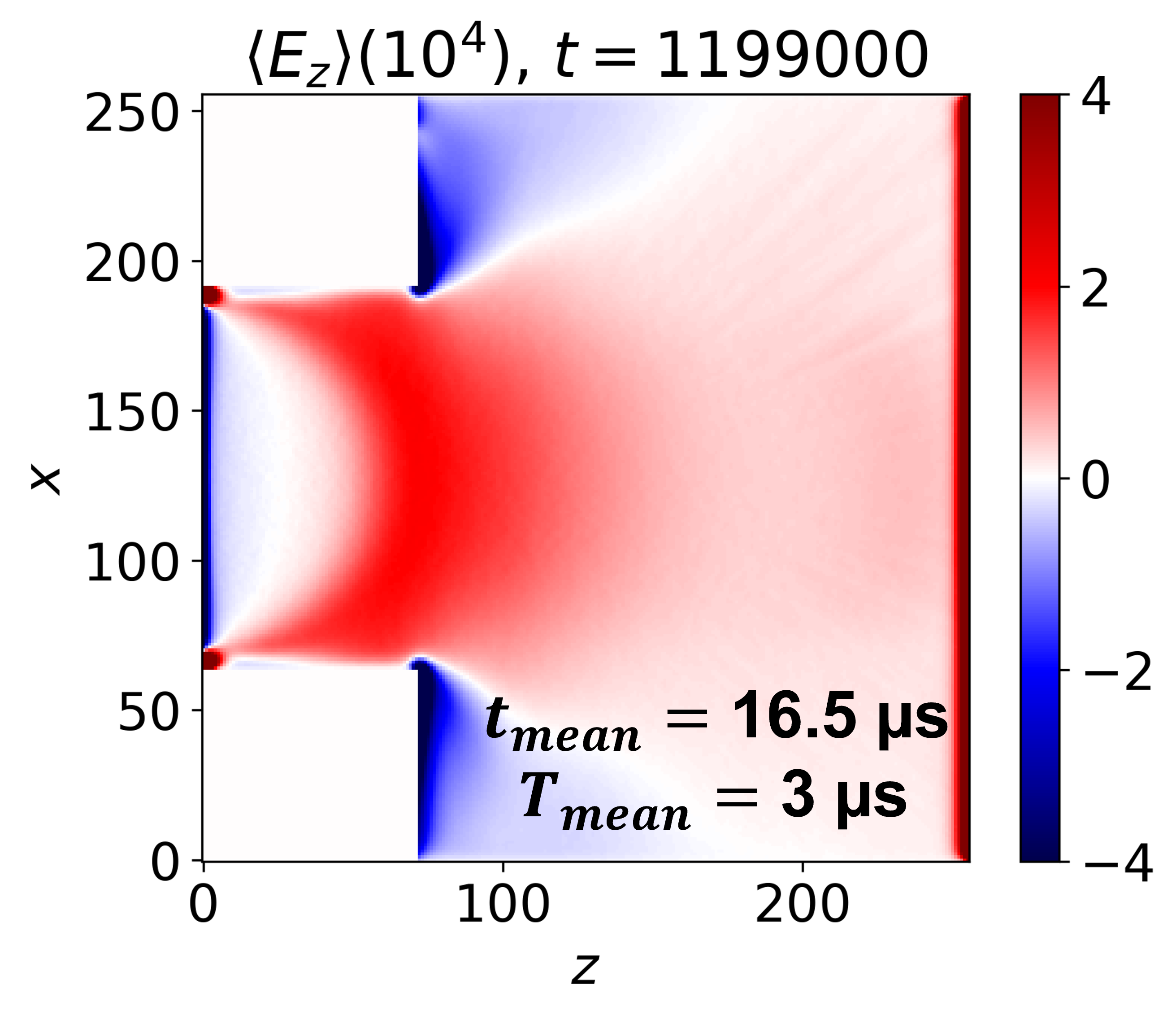}
\includegraphics[width=0.3\textwidth]{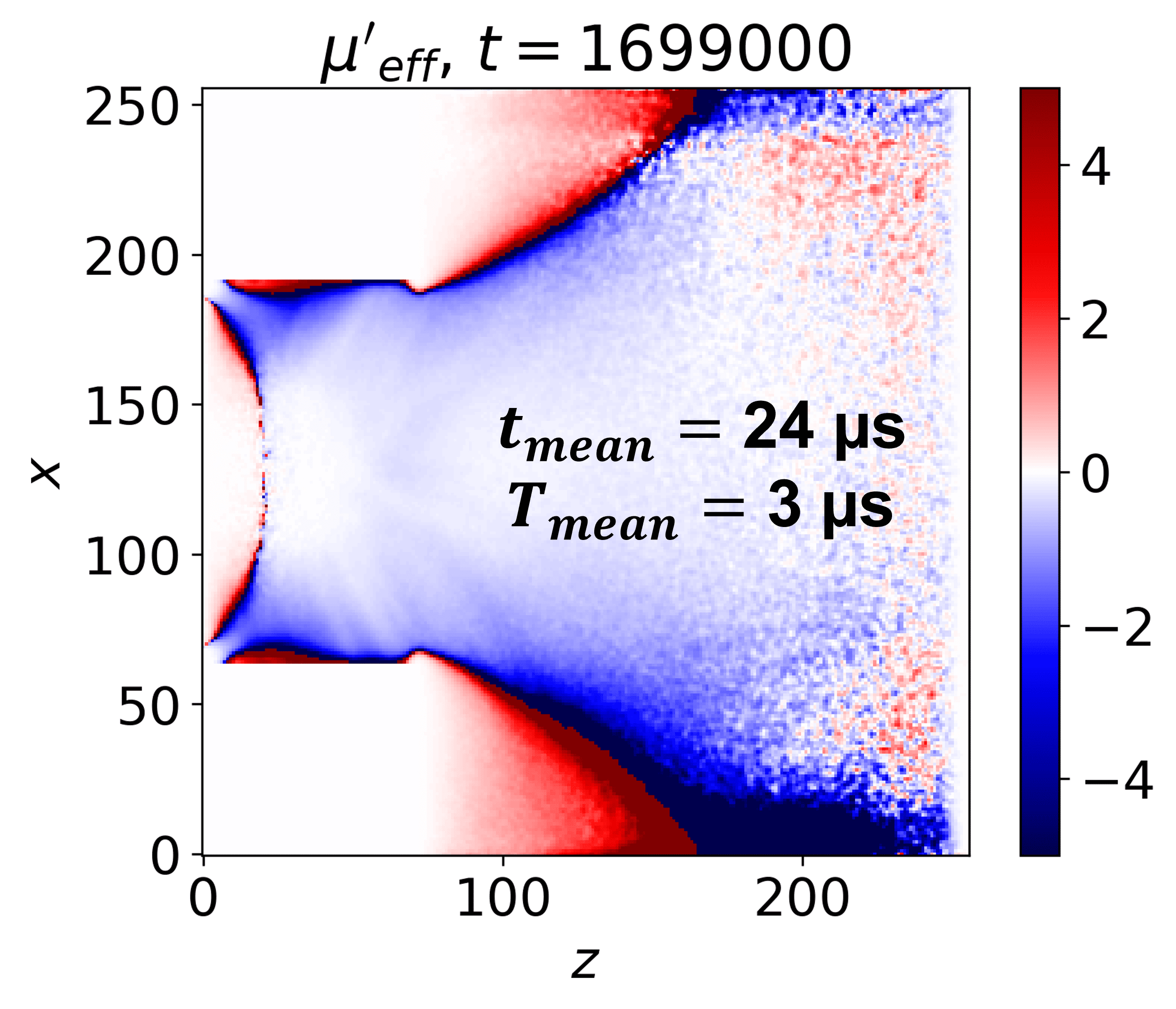}
\includegraphics[width=0.3\textwidth]{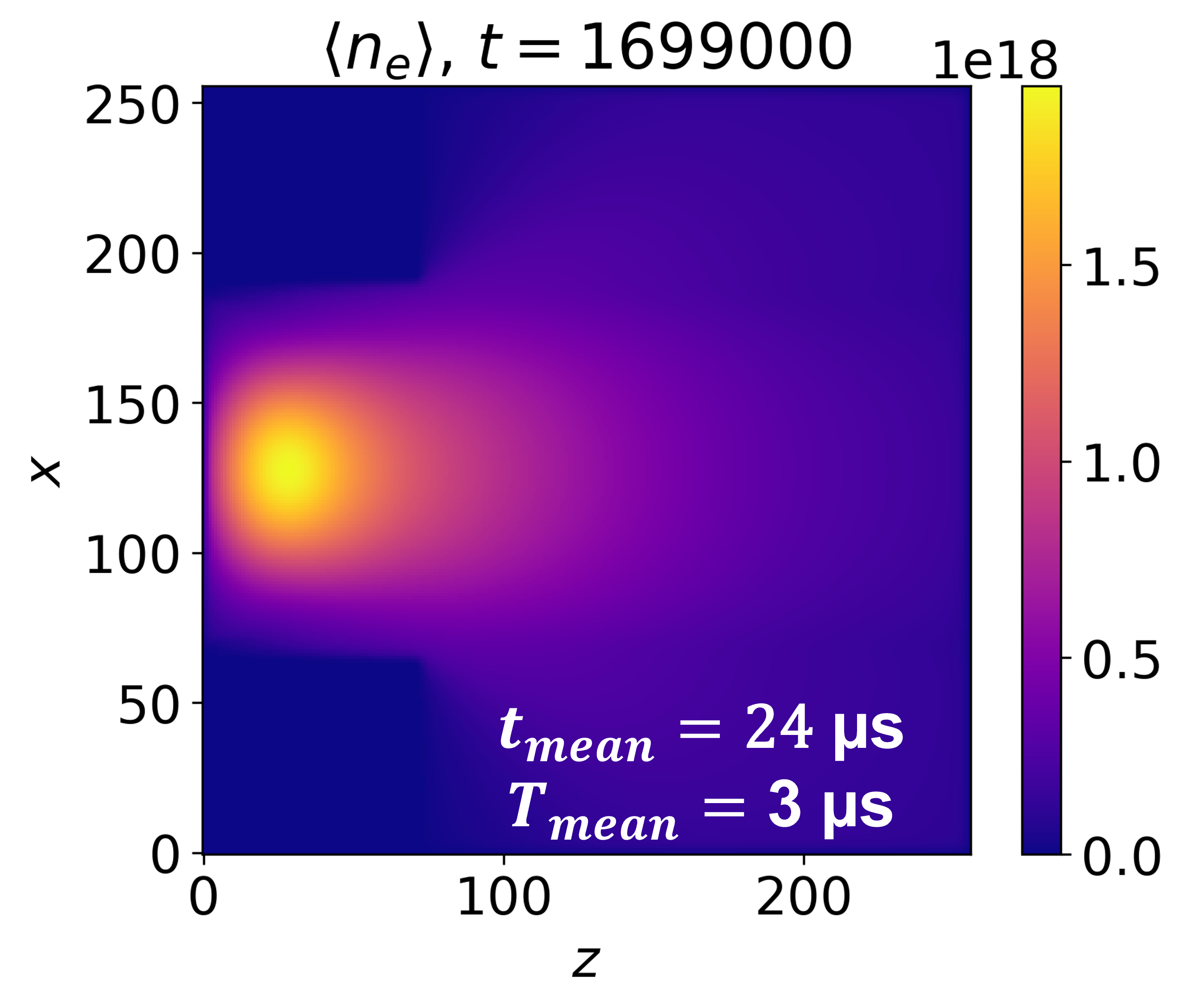}
\includegraphics[width=0.3\textwidth]{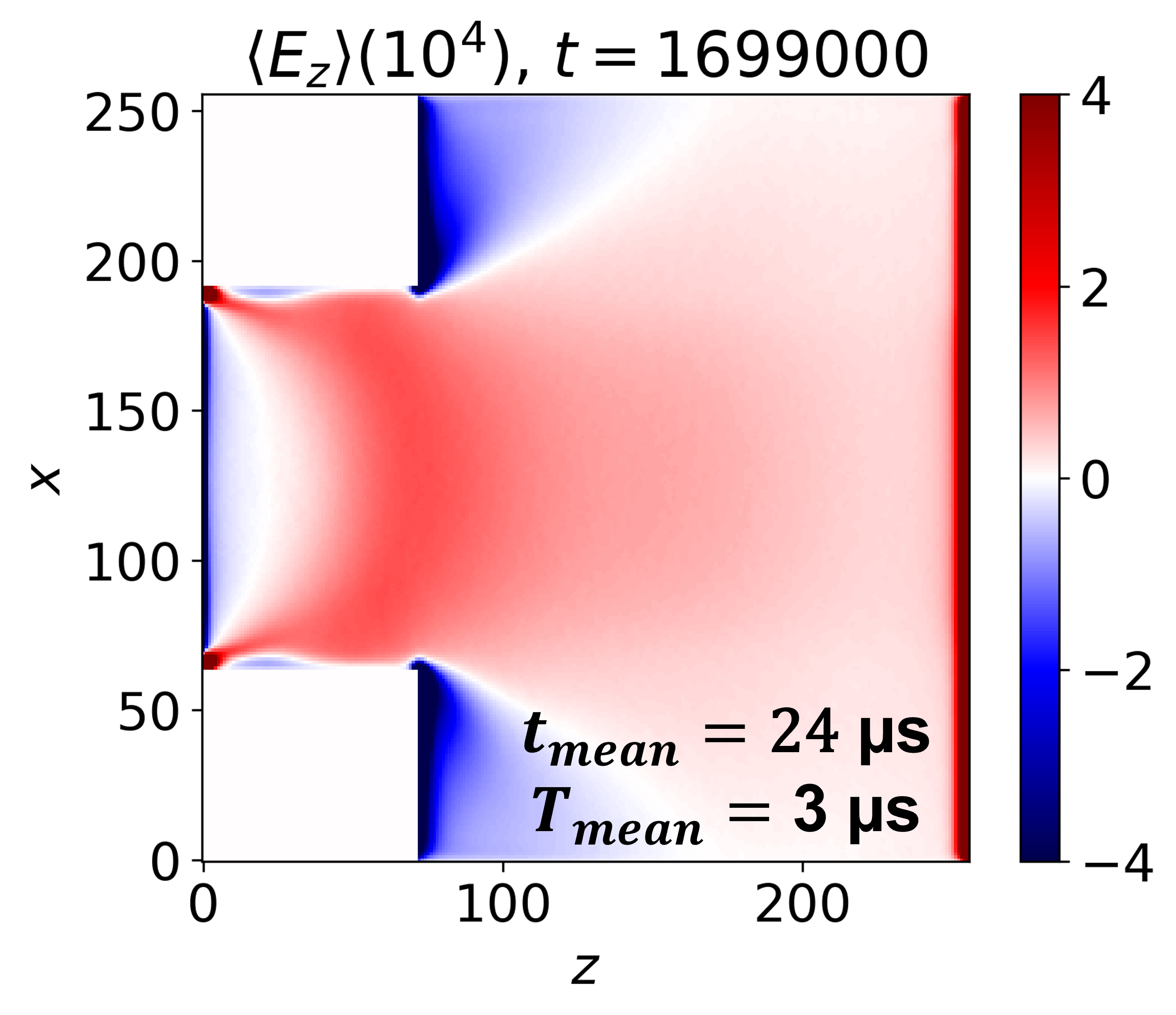}
\caption{
The effective electron mobility, averaged electron density, averaged axial electric field
of the Analytic-B case
evaluated using different time periods.
}
\label{fig:mu_1199000_03}
\end{figure}

\begin{figure}[ht]
\centering
\includegraphics[width=0.3\textwidth]{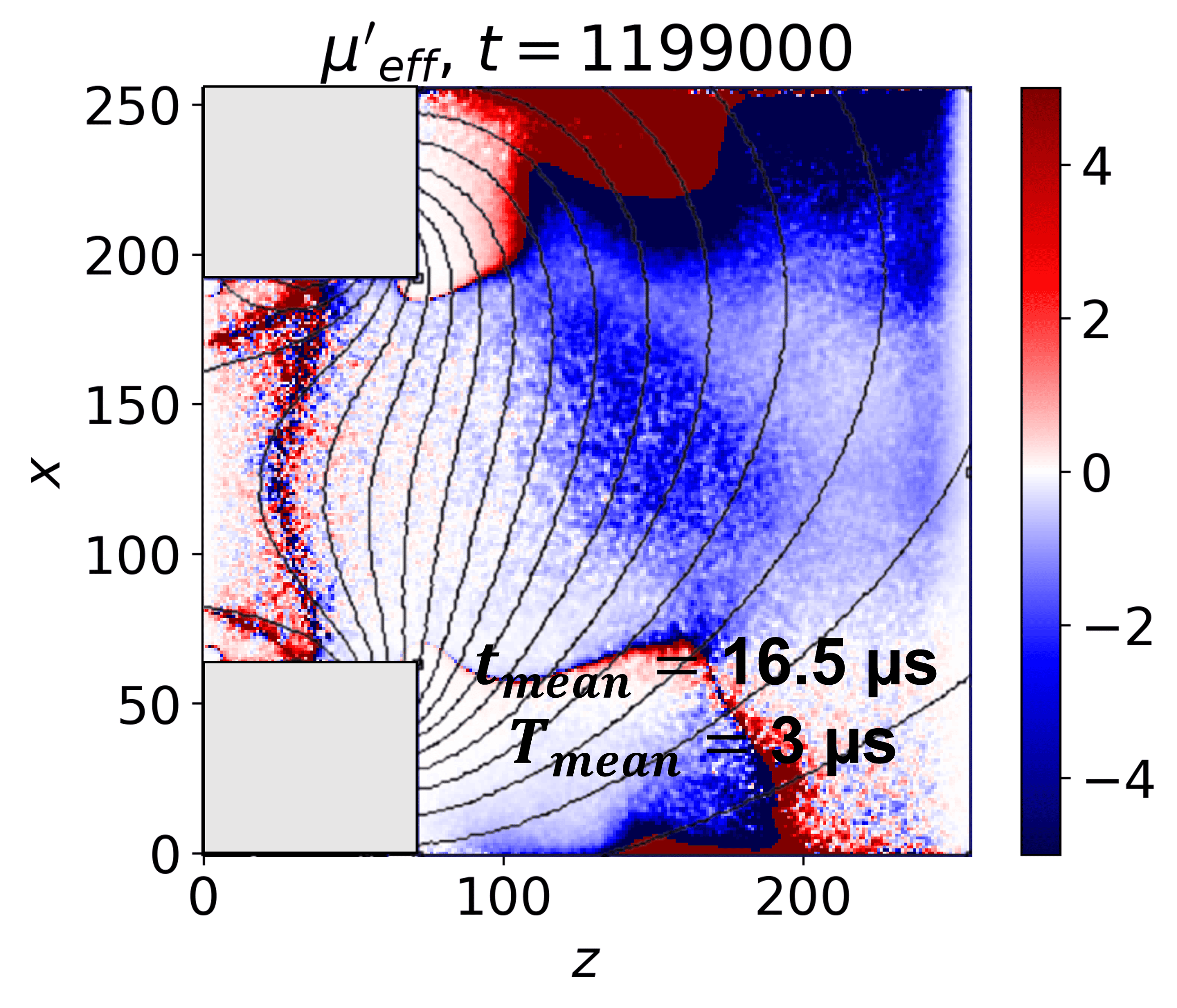}
\includegraphics[width=0.3\textwidth]{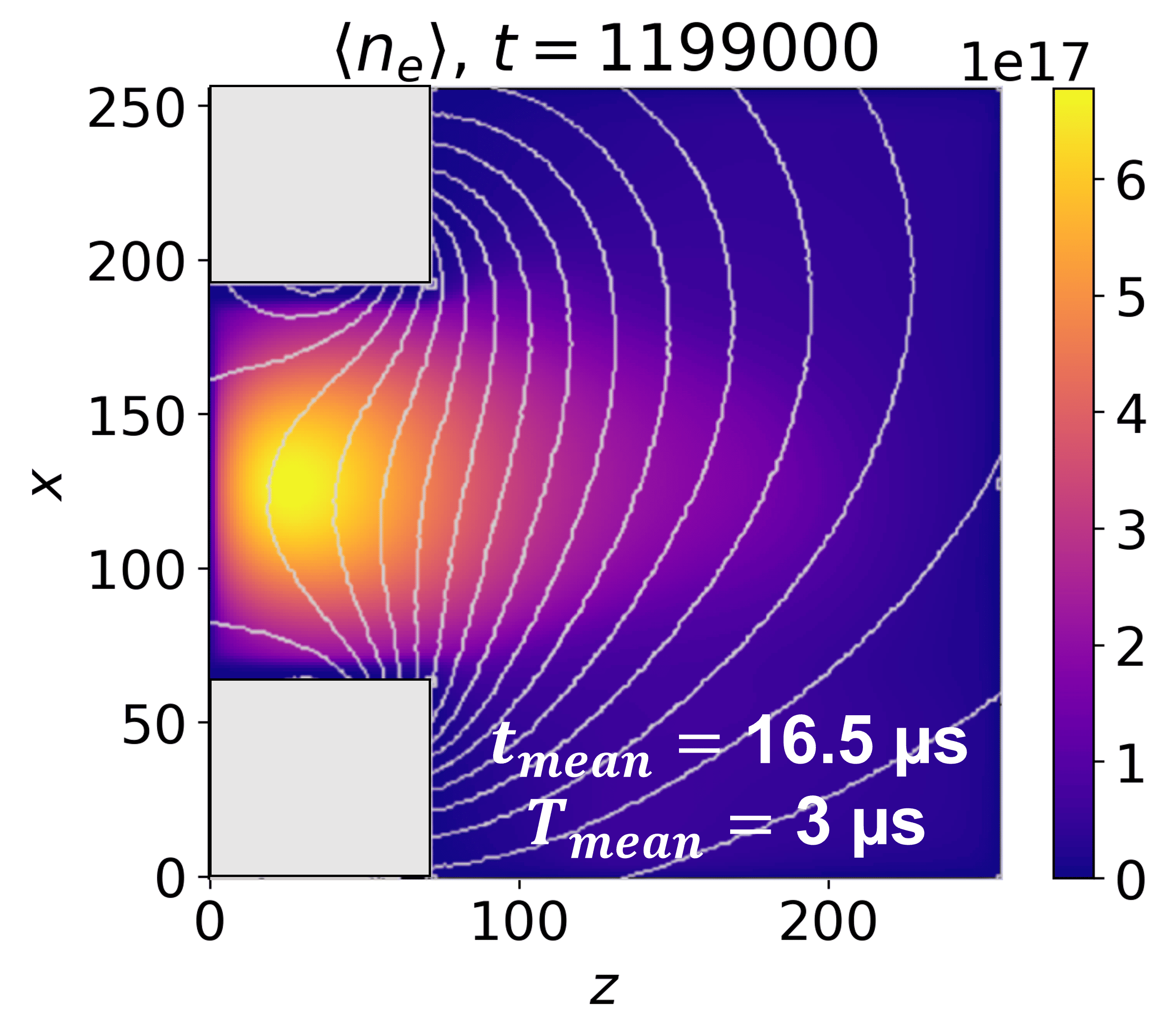}
\includegraphics[width=0.3\textwidth]{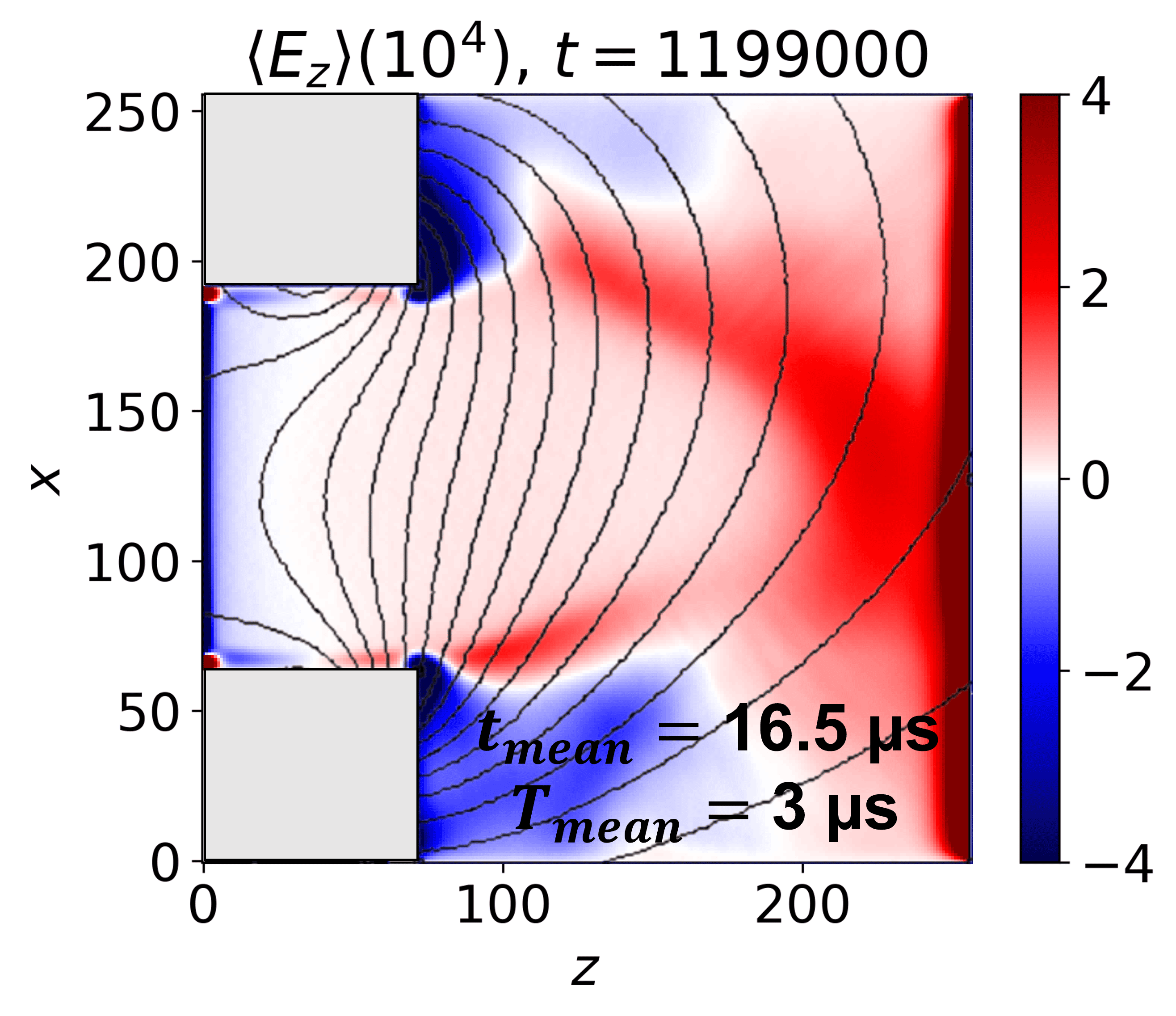}
\includegraphics[width=0.3\textwidth]{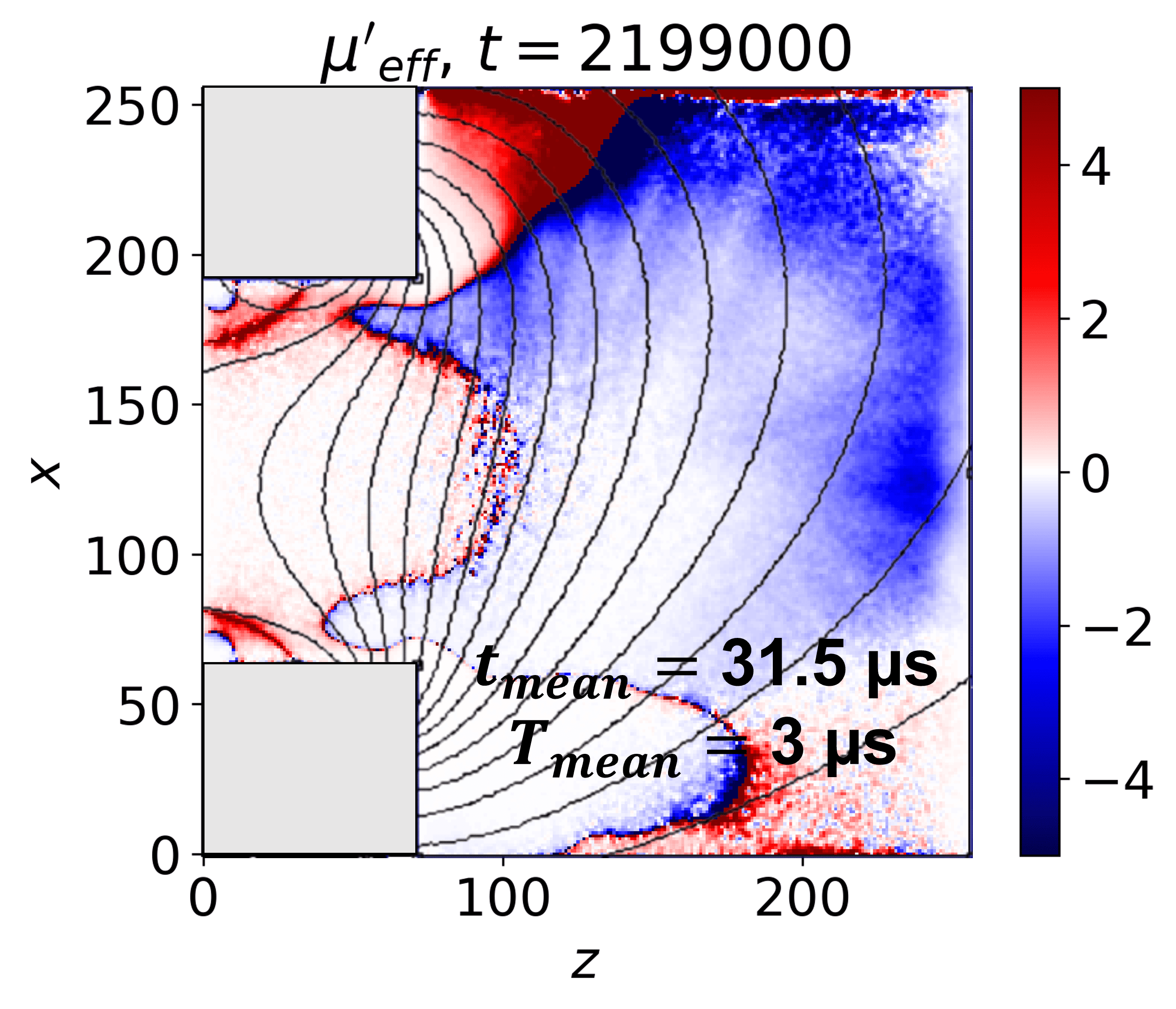}
\includegraphics[width=0.3\textwidth]{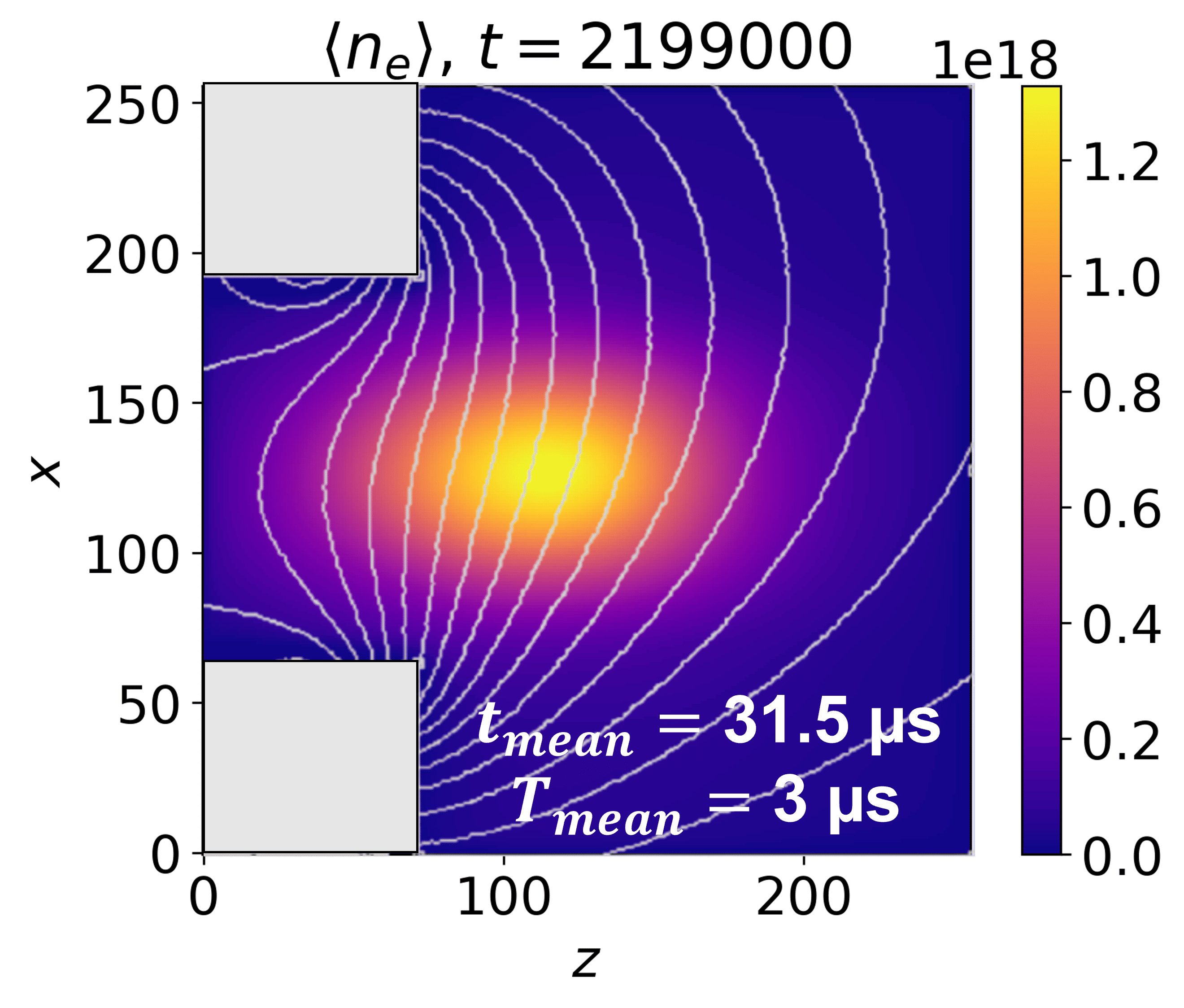}
\includegraphics[width=0.3\textwidth]{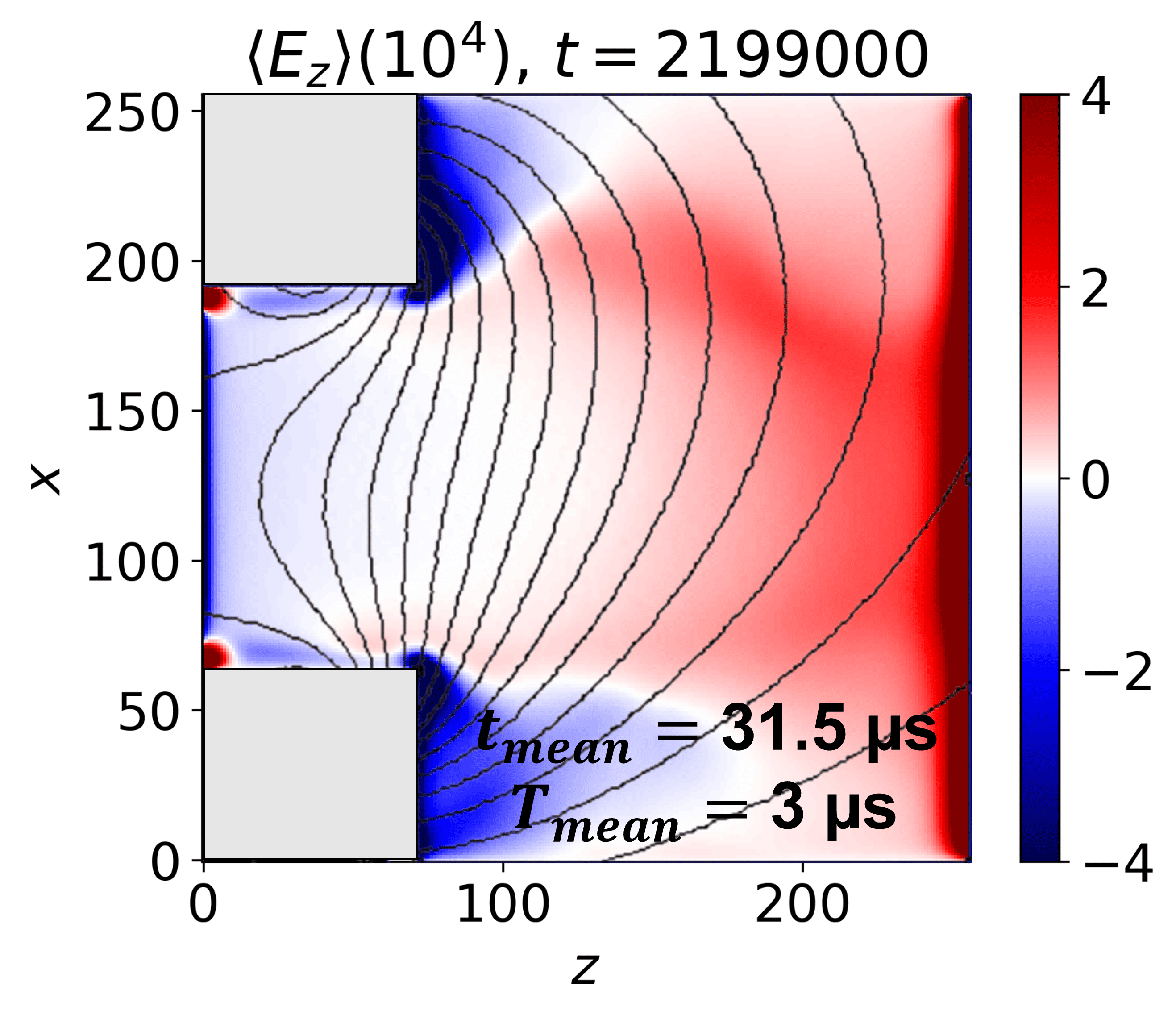}
\caption{
The effective electron mobility, averaged electron density, averaged axial electric field
of the Weak-B case
evaluated using different time periods.
}
\label{fig:mu_1199000_04}
\end{figure}

\begin{figure}[!b]
\centering
\includegraphics[width=0.24\textwidth]{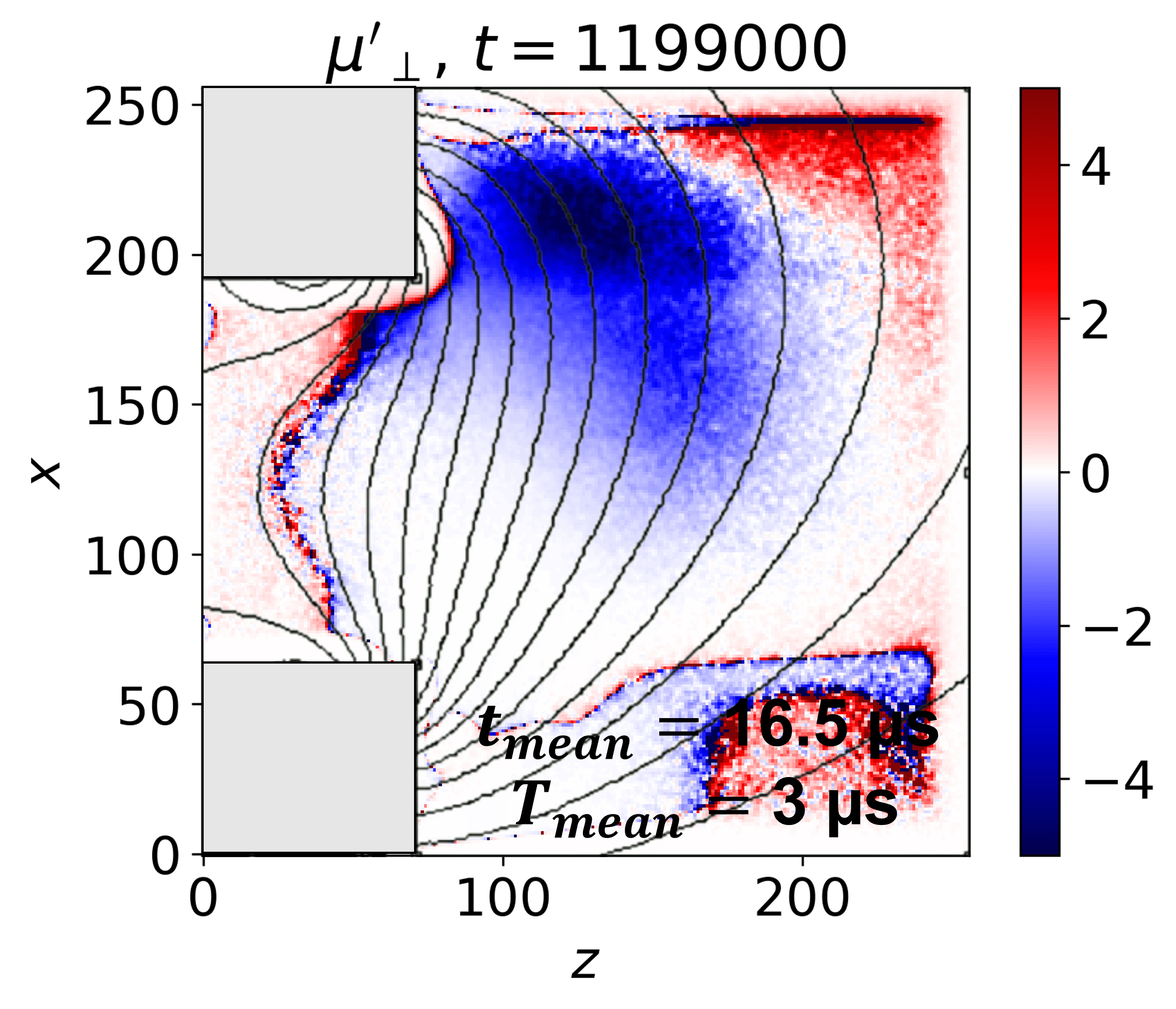}
\includegraphics[width=0.24\textwidth]{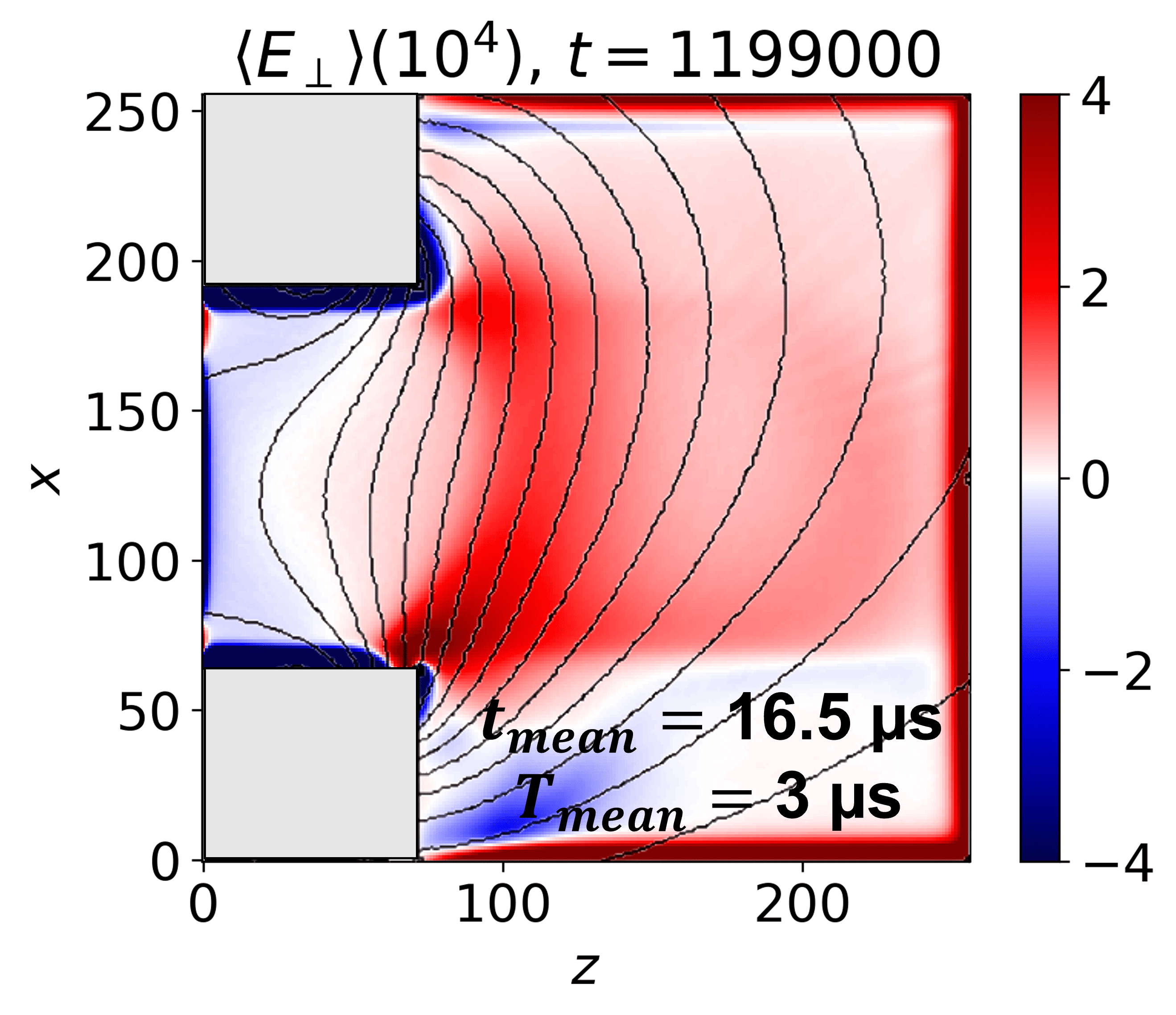}
\includegraphics[width=0.24\textwidth]{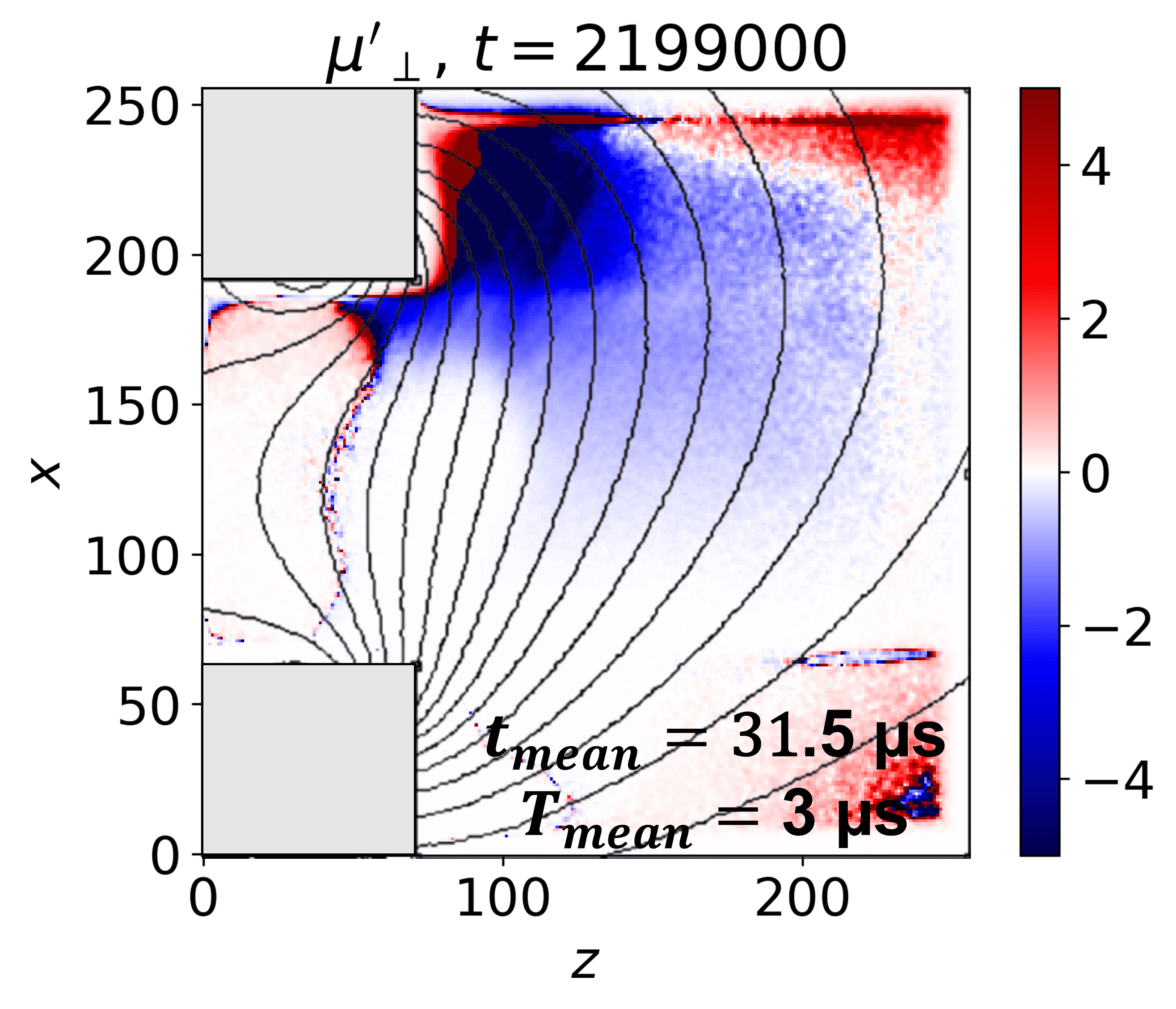}
\includegraphics[width=0.24\textwidth]{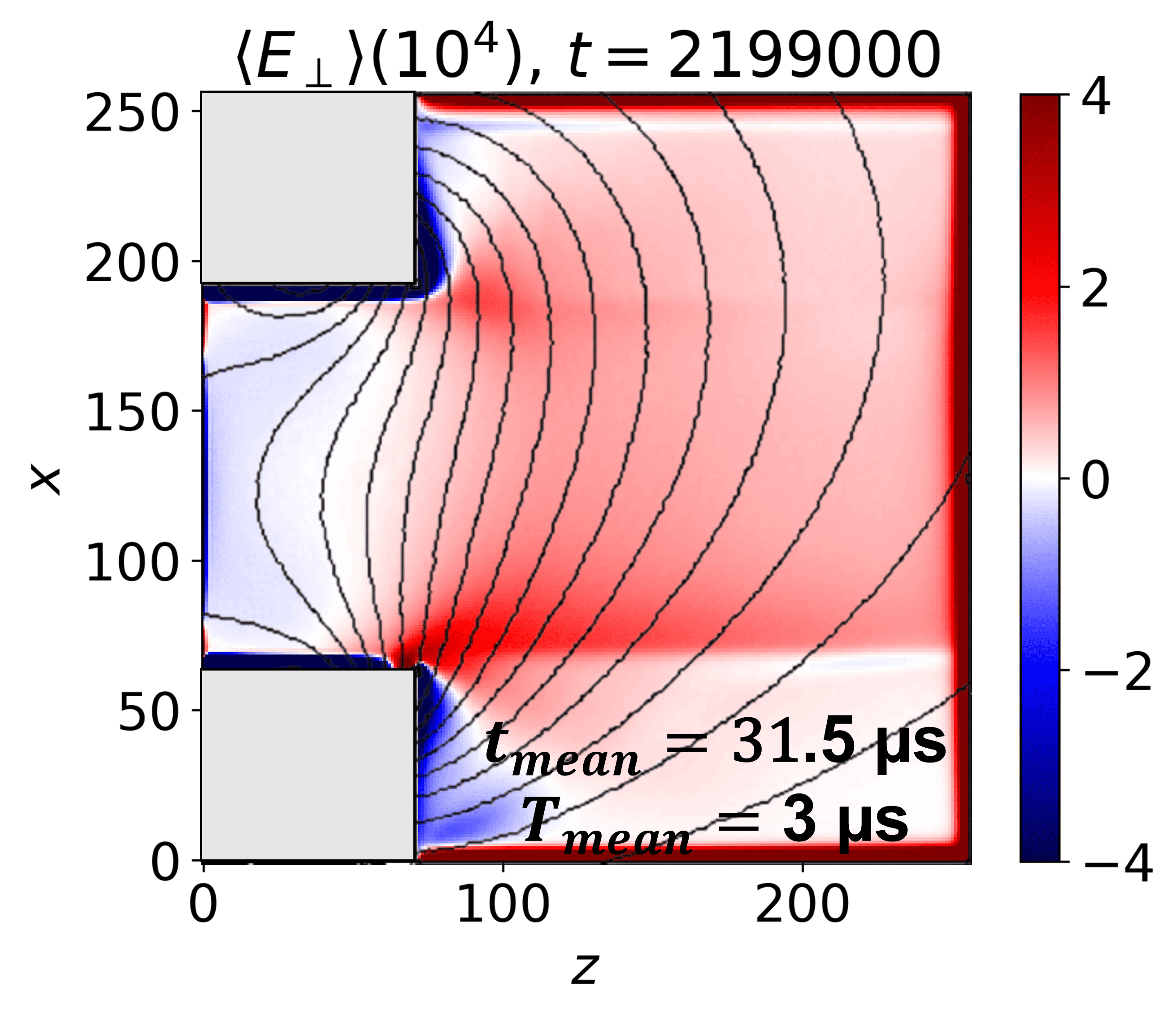}
\caption{
The perpendicular electron mobility and the averaged perpendicular electric field
of the Strong-B case
evaluated using different time periods.
}
\label{fig:mup}
\end{figure}

\begin{figure}[ht]
\centering
\includegraphics[width=0.3\textwidth]{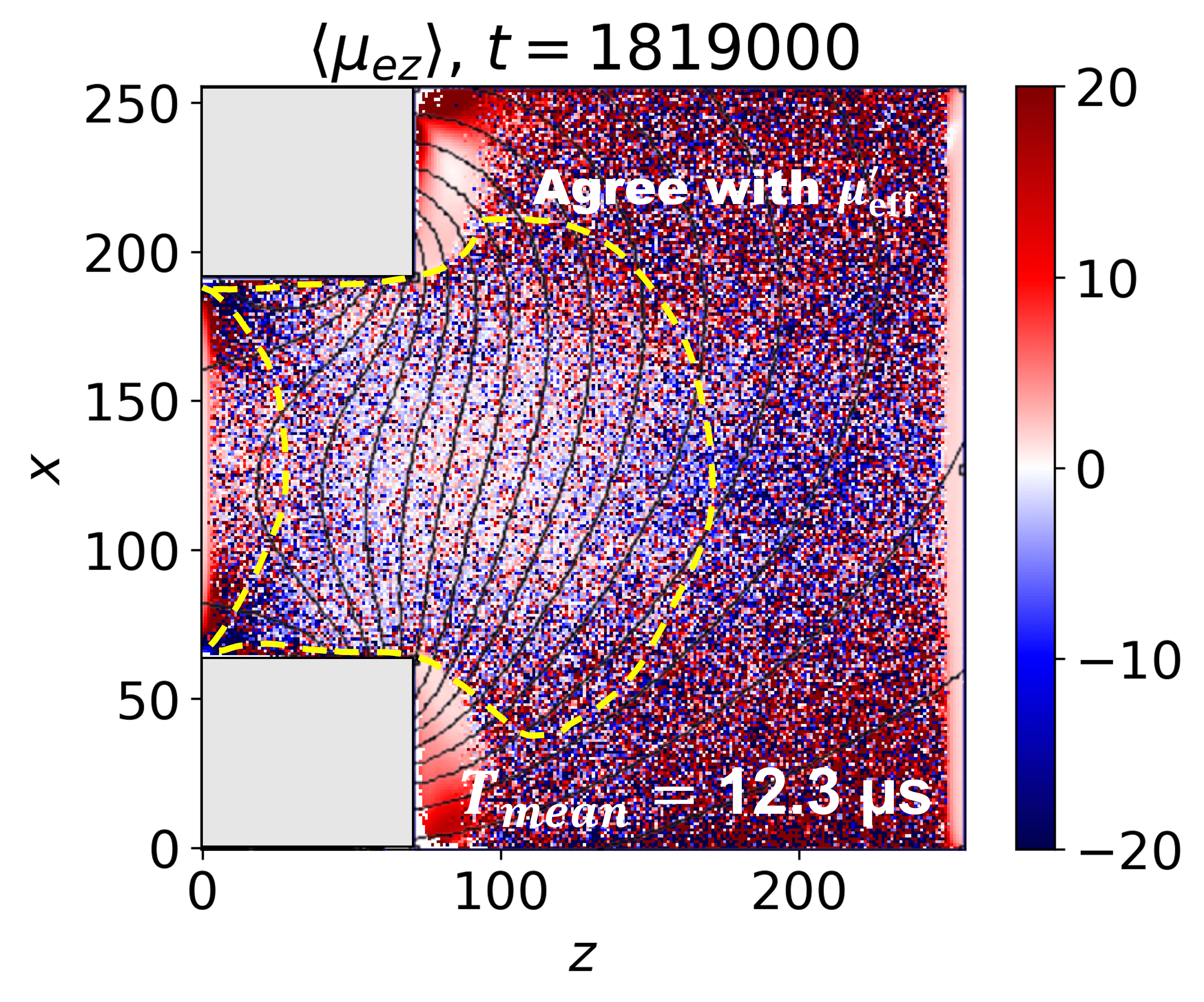}
\includegraphics[width=0.3\textwidth]{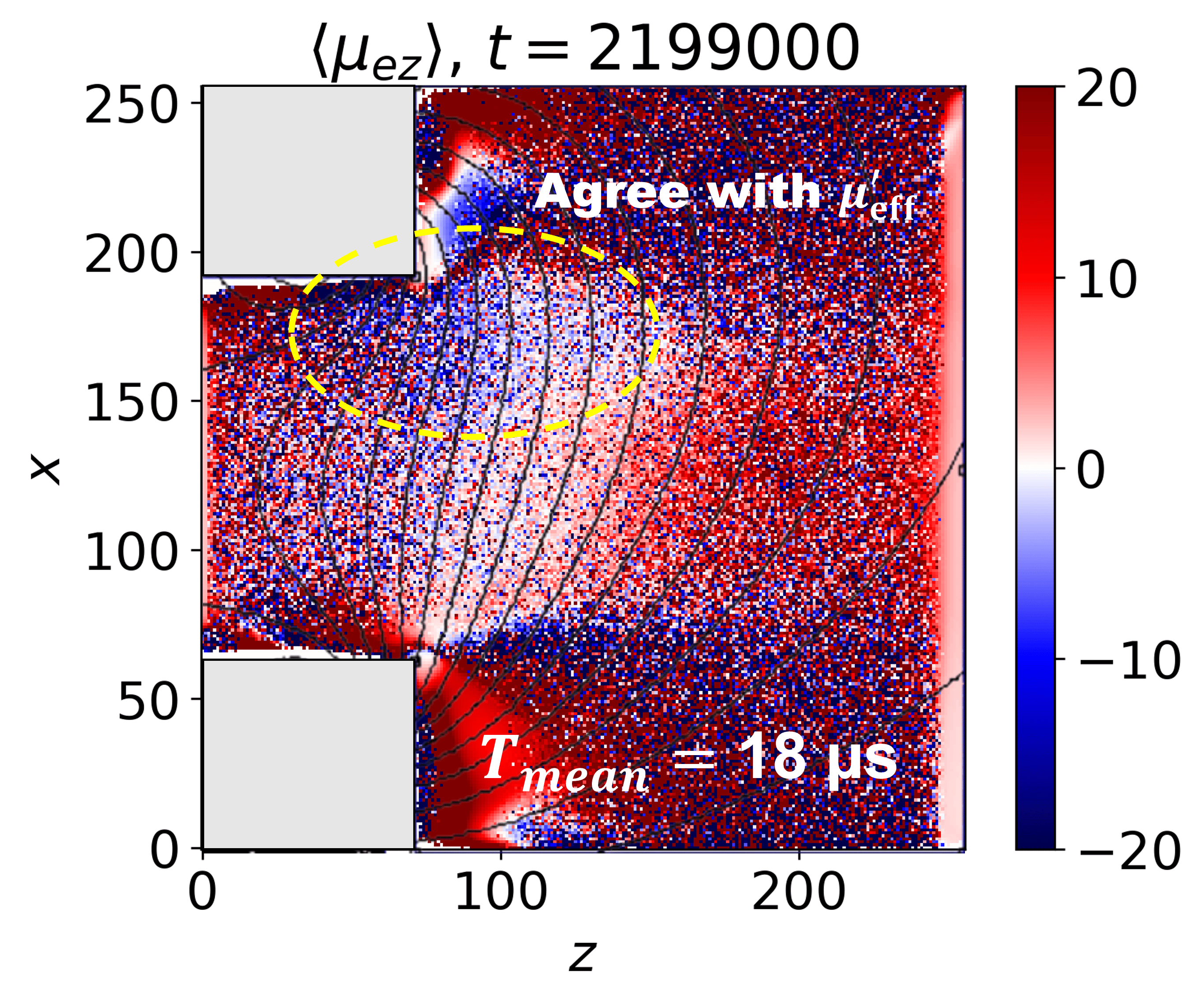}
\includegraphics[width=0.3\textwidth]{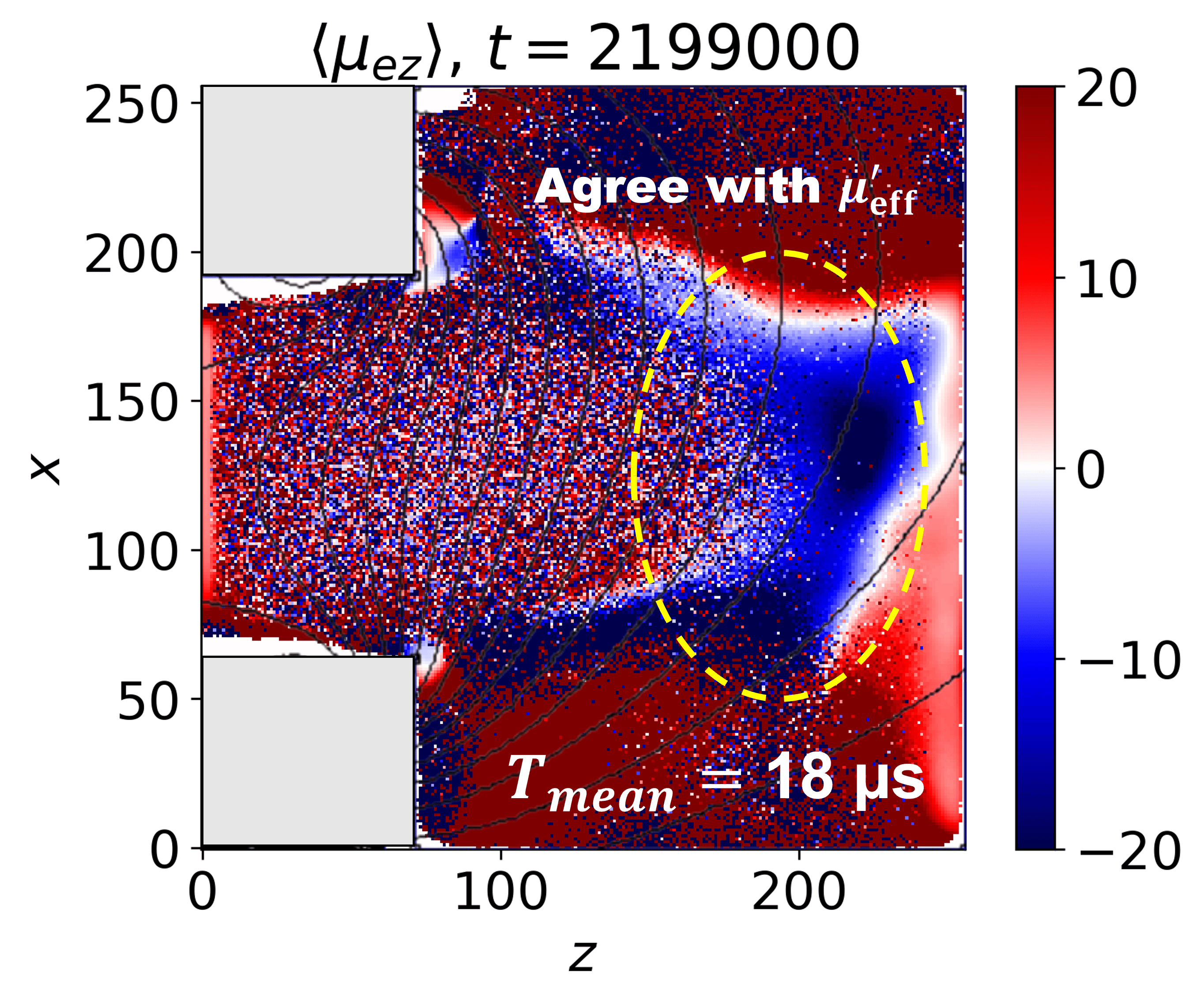}
\caption{
The electron mobility computed using
the mean electron axial velocity and
the axial electric field in PIC simulations
for the case Analytic-B (left),
Strong-B (middle)
and Weak-B (right).
}
\label{fig:mupic}
\end{figure}

\subsection{Correlation between
the electron density and the azimuthal electric field}

The correlation terms between the
electron density and the azimuthal electric
field,
i.e., $\left< n_e E_y \right>$,
averaged spatially along $y$
and temporally within different time periods,
are presented in Fig.\ref{fig:nE_01}
for the Strong-B case.
As derived by T. Lafleur et al.
\cite{Lafleur1},
the correlation term contributes to the electron
mobility as
\cite{IEPC-2025-063}
\begin{equation}\label{eq:mueffp}
    \mu_{\text{eff}}' \equiv
    - \dfrac{\left< n_e E_y \right>}{B_x n_e E_z},
\end{equation}
ignoring the classical transport due
to electron-neutral collisions.
Note that $\mu_{\text{eff}}'$
is defined as negative for electron
transport to the anode in this paper.
The first three panels share the same color bar range,
and are averaged around $t_{mean}=16.5$,
19.5, and 22.5 $\mu$s, respectively,
within time length $T_{mean} = 3$ $\mu$s
using data at every 1000 time steps (namely 15 ns).
One can find these three corresponding times in
Fig.\ref{fig:ni2} (the Strong-B case line)
around the local minimal of $n_i$,
such that the magnitude of $\left< n_e E_y \right>$
decreases from $5 \times 10^{19}$ V/m$^{4}$ at
$t_{mean}=16.5$ $\mu$s
to a local minimal about
$2 \times 10^{19}$ V/m$^{4}$
at $t_{mean}=19.5$ $\mu$s
and increases back at $t_{mean}=22.5$ $\mu$s.
Since $\left< n_e E_y \right>$ reflects the
electron transport due to
the field oscillations of the instability,
these three panels indicate that
the major electron transport occurs
at the upper region in the radial direction
where the magnetic field is relative low
compared to the lower radial region.
In the axial direction,
the major electron transport due to the instability
occurs at the thruster exit
and the plume region near the exit.

Then, the other three panels
in Fig.\ref{fig:nE_01}
share the same color bar range,
and are averaged around
$t_{mean}=26.4$,
28.5, and 31.5 $\mu$s, respectively.
These time periods correspond to
the peak of $n_i$ as shown in Fig.\ref{fig:ni2},
thus the magnitude of $\left< n_e E_y \right>$
increases over time.
Compare the results,
it can be seen that not only the magnitude,
but also the shape of the electron transport path
changes
for different stages during
the low-frequency oscillating of $n_i$
due to the breathing mode ionization.
At low $n_i$, the electron transport path
is more spread, while
at high $n_i$, it is more focused.

Next, the corresponding $\left< n_e E_y \right>$
results of
the Analytic-B case are shown in Fig.\ref{fig:nE_03}.
Again, the considered three times
($t_{mean}=$16.5 $\mu$s, 19.425 $\mu$s, and 24 $\mu$s)
correspond to a gradual increase of the
$n_i$ as shown in Fig.\ref{fig:ni2}.
Besides the detailed difference
of the results at these three times,
the overall trends are the same,
namely high electron transport due to the
instability are expected to occur
at the thruster exit and near the surface
inside the discharge channel,
which is greatly different from that of
the Strong-B case.
This result indicates that
the commonly used analytic B field considering
only the radial component should be replaced
by more realistic B field 
to obtain more accurate electron transport.

At last in this subsection,
the corresponding $\left< n_e E_y \right>$
results of the Weak-B case are shown in Fig.\ref{fig:nE_04}.
Many results at different time steps are presented
during the growth of $n_i$,
but overall the electron transport looks very
different from the Strong-B and the Analytic-B
case.
The results at $t_{mean}=$16.5 $\mu$s
and 31.5 $\mu$s are closer,
because at least $n_i$ are not too low,
and the EDI could be established and evolve.
Nevertheless, as can be seen in Fig.\ref{fig:ni2},
in between these two times,
$n_i$ drops to and maintains at a very low value
(about $0.4 \times 10^{17}$ m$^{-3}$),
thus the EDI could not be established,
and the plasma behavior is completely abnormal.
Since the Weak-B case has only a peak $B$
value about 100 Gauss along the center line,
it could be expected that electrons are not well
confined by the magnetic field
compared to the other two cases.

% \begin{figure}[ht]
% \centering

% \caption{
% The correlation between the electron density and the azimuthal
% electric field of the Analytic-B case
% evaluated using different time periods.
% }
% \label{fig:nE_01}
% \end{figure}

\subsection{The instability induced effective electron mobility}

In this subsection,
using the $\left< n_e E_y \right>$ correlation
obtained in the PIC simulations,
the electron mobility due to the instability
can be computed according to Eq.(\ref{eq:mueffp}).
During the computation,
since the averaged
electron density and the axial electric field
are used,
they are plotted alongside with $\mu_\textrm{eff}'$
at typical times
for the Strong-B case in Fig.\ref{fig:mu_1199000},
the Analytic-B case in Fig.\ref{fig:mu_1199000_03},
and the Weak-B case in Fig.\ref{fig:mu_1199000_04}.

For the Strong-B case in Fig.\ref{fig:mu_1199000},
two typical times are chosen,
namely 16.5 $\mu$s and 31.5 $\mu$s.
Comparing $\mu'_{\textrm{eff}}$ to
the corresponding $\left< n_e E_y \right>$
shown in Fig.\ref{fig:nE_01},
high values of $\left< n_e E_y \right>$
lead to high values of $\mu'_{\textrm{eff}}$ as expected,
except for a dividing line between
negative and positive $\mu'_{\textrm{eff}}$
in the upper radial region.
It can be seen later that
this dividing line can be eliminated
by evaluating the electron mobility
perpendicular to the magnetic field lines,
but we still prefer to first present
$\mu'_{\textrm{eff}}$ defined along $z$.
In addition, comparing the
axial electric field profiles at
these two different time steps shown in
Fig.\ref{fig:mu_1199000},
it can be seen that the strength of $E_z$
at the exit drops when $n_i$ is near the peak value
at 31.5 $\mu$s
compared with that at 16.5 $\mu$s.

Next, looking at the $\mu'_{\textrm{eff}}$ of the Analytic-B case
in Fig.\ref{fig:mu_1199000_03},
$t_{mean}=16.5$ and 24 $\mu$s
are chosen,
the corresponding $\left< n_e E_y \right>$ plots
are presented in Fig.\ref{fig:nE_01}.
It can be seen that near the surface
inside the discharge channel,
relatively large $\mu'_{\textrm{eff}}$ is expected,
but the high values of $\left< n_e E_y \right>$
in the center at the thruster exit
lead to reduced extent of $\mu'_{\textrm{eff}}$,
due to the combined effects of $n_e$, $E_y$,
and $B_x$.

The $\mu'_{\textrm{eff}}$ of the Weak-B case is shown
in Fig.\ref{fig:mu_1199000_04}.
Again, higher values of $\left< n_e E_y \right>$
(shown in Fig.\ref{fig:nE_04})
lead to stronger electron mobility to the anode.
Because the magnetic field is not strong enough in this case,
the peak electron density moves from the region inside the channel
at 16.5 $\mu$s
to the region outside of the channel at 31.5 $\mu$s.
The axial electric field can not be established near the thruster
exit,
but rather moves to the down-stream region,
indicating an abnormal Hall thruster discharge.

In addition,
because Eq.(\ref{eq:mueffp})
was derived originally based on the purely radial
magnetic field assumption,
but in the Strong-B and the Weak-B case,
the magnetic field
contains both radial and axial components.
It would be more reasonable to
analyze the electron mobility perpendicular
to the $B$ field, i.e.,
\begin{equation}\label{eq:mueffpp}
    \mu_{\perp}' \equiv
    - \dfrac{\left< n_e E_y \right>}{B n_e E_\perp},
\end{equation}
where $B=(B_x^2 + B_z^2)^{1/2}$,
and $E_\perp$ can be evaluated
by transforming $\bm{E}=(E_\parallel,E_\perp)$ into a new coordinate system
defined by the two axes that are
parallel and perpendicular to the local $\bm{B}$,
\begin{equation}
    E_\perp = \dfrac{B_z E_x - B_x E_z}{B}.
\end{equation}
The perpendicular electron mobility for the Strong-B case is plotted
in Fig.\ref{fig:mup}.
As can be seen compared with the normal $\mu'_{\textrm{eff}}$
plot of the Strong-B case in Fig.\ref{fig:mu_1199000},
the dividing line between
negative and positive $\mu'_{\textrm{eff}}$
in the upper radial region becomes less pronounced
under the diagnosis of perpendicular electron mobility.
Thus, $\mu_{\perp}'$ would be a better choice for future
analyses of the electron mobility
when the magnetic field contains both
radial and axial components.

\subsection{The PIC electron mobility}

The electron mobility in PIC simulations
can be directly evaluated by $\mu_{\textrm{ez}}$,
\begin{equation}
    \mu_{\textrm{ez}} \equiv \dfrac{u_{ez}}{E_z},
\end{equation}
where $u_{ez}$ denotes the mean electron axial velocity and
$E_z$ is the axial electric field.
Because it is found that the noise is large,
longer periods are used to obtain
averaged results over time
and space in $y$,
indicated by $\left< \mu_{ez} \right>$.
The results of the three cases
with labeled averaging periods $T_{\textrm{mean}}$
are shown in Fig.\ref{fig:mupic}.

The $\left< \mu_{ez} \right>$
of the Analytic-B case can be compared to
the $\mu'_{\textrm{eff}}$ shown in
Fig.\ref{fig:mu_1199000_03}.
Although $\left< \mu_{ez} \right>$ is noisy,
more negative blue points inside
the channel near the surfaces and
in the plume near the exit indicate
the same trend and shape that can be
seen in $\mu'_{\textrm{eff}}$ as well,
which is circled by the dashed yellow line
in Fig.\ref{fig:mupic}.
Similarly, the middle plot of
$\left< \mu_{ez} \right>$ of the Strong-B case
can be compared to the $\mu'_{\textrm{eff}}$
shown in Fig.\ref{fig:mu_1199000}.
Again, similar trends can be seen,
indicating that $\left< \mu_{ez} \right>$ agrees
with $\mu'_{\textrm{eff}}$.
At last, $\left< \mu_{ez} \right>$ of the Weak-B
case can be compared to $\mu'_{\textrm{eff}}$
shown in Fig.\ref{fig:mu_1199000_04}.
Negative electron transport regions can be
identified in the downstream plume region,
which agrees with $\mu'_{\textrm{eff}}$.

\subsection{The inverse Hall parameter}

\begin{figure}[ht]
\centering
(a)
\includegraphics[width=0.45\textwidth]{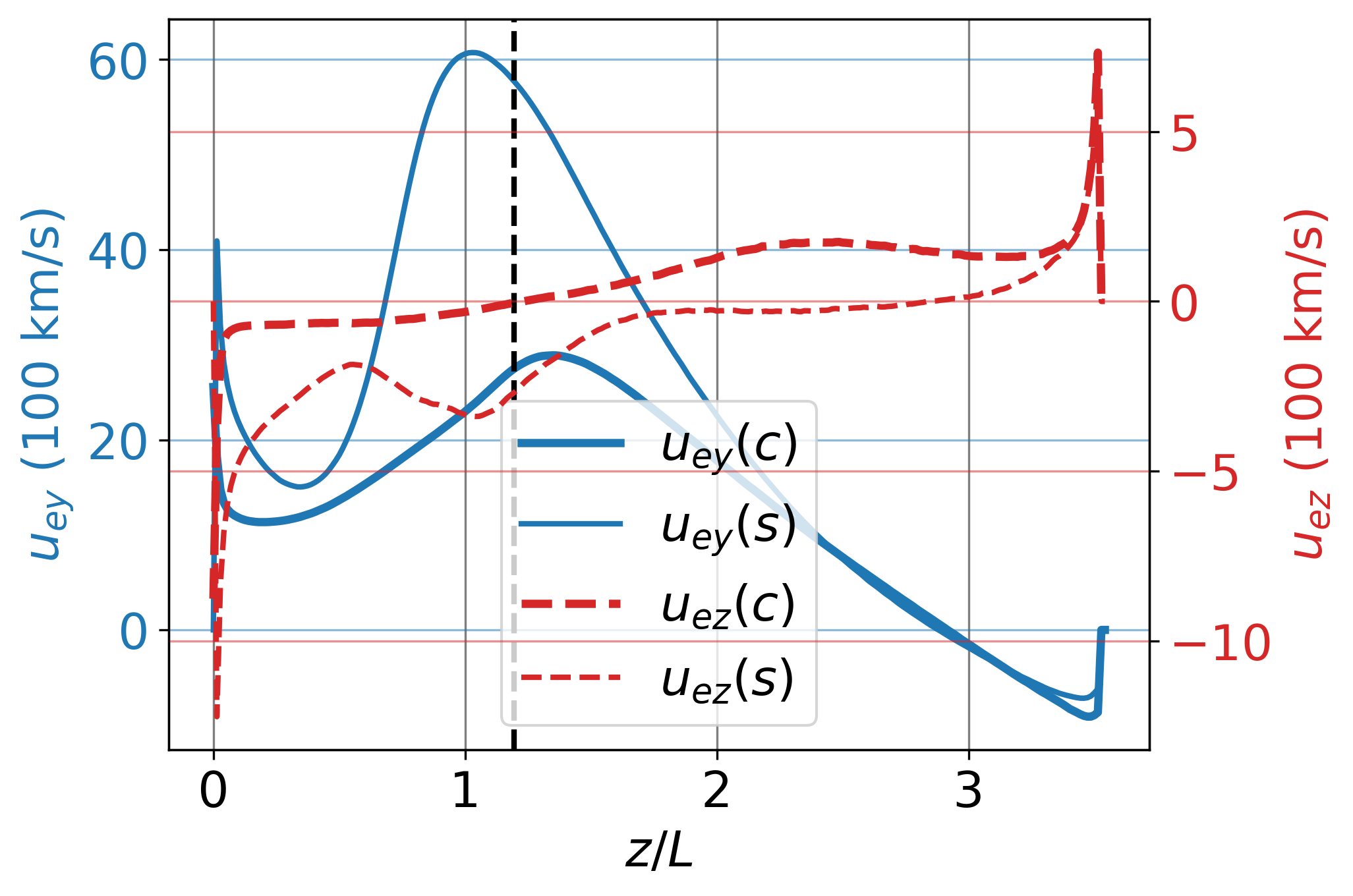}
(b)
\includegraphics[width=0.4\textwidth]{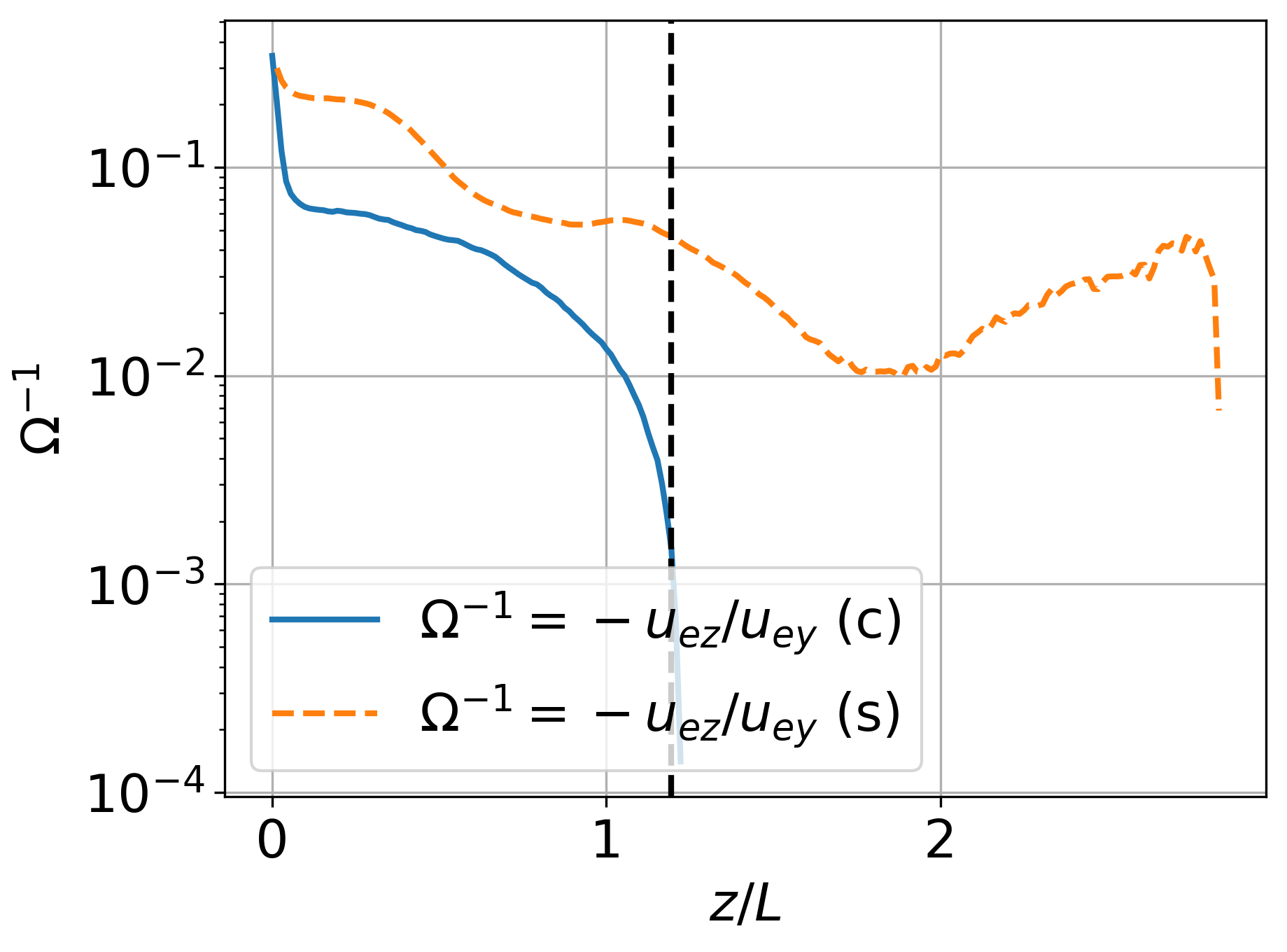}
\caption{
(a) Mean azimuthal ($u_{ey}$) and axial 
($u_{ez}$) electron velocity profiles
over $z$ along the centerline labeled ``c''
or near the surface of the upper radial region
labeled by ``s''.
(b) The computed inverse Hall parameter
over $z$.
$L=0.72$ cm denotes the channel length,
and the maximum B field
on the cneterline at about 0.86 cm is indicated using
black dashed lines.
}
\label{fig:omg}
\end{figure}

At last, motivated by
P. Roberts and B. Jorns
\cite{PhysRevLett.132.135301},
the inverse Hall parameter is plotted
for the Strong-B case.
The mean azimuthal ($u_{ey}$) and axial 
($u_{ez}$) electron velocity profiles
over $z$ along the centerline
or near the surface of the upper radial region (10 cells
away from the surface)
are given in Fig.\ref{fig:omg} (a).
These velocities are averaged over
the azimuthal space and time period of
12.3 $\mu$s as the same as
that in Fig.\ref{fig:mupic}
of the Strong-B case.
As we can see $u_{ey}$ increases until near the
maximum B field, then drops;
$u_{ez}$ increases to zero
between $z/L =1$ and 2,
where $L=L_{zc}=0.72$ cm is the channel length.
When computing the inverse Hall parameter
$\Omega^{-1}=-u_{ez}/u_{ey}$,
only the ranges with $u_{ez}<0$
are considered, such that
$\Omega^{-1}>0$.
The two $\Omega^{-1}>0$ curves
on the centerline and near the channel surface
are plotted in Fig.\ref{fig:omg} (b).
Because the magnetic field configuration,
the maximum B field location compared
to the channel exit,
and many other parameters
are all different from those applied
in \cite{PhysRevLett.132.135301},
no agreement is expected,
but reasonable magnitudes of $\Omega^{-1}$
are achieved.
In addition,
it is shown that the axial distribution of 
$\Omega^{-1}$ can vary significantly across different radial regions, particularly when considering the perspective of distinct electron transport paths in various radial zones as identified in this study.

\section{Discussions and conclusions}
\label{sec:conclusion}

The most important results obtained in this
paper are those 2D radial-axial profiles
of the correlation term $\left< n_e E_y \right>$ or the derived
electron effective mobility $\mu'_{\textrm{eff}}$
(or $\mu'_{\perp}$) due to the instabilities.
They can only be obtained from fully 3D PIC simulations
by averaging over space in the azimuthal direction
and time. It has been shown in this paper
that different magnetic field configurations
and strengths lead to different 2D profiles
of electron mobility.
This indicates a pathway of using the
3D PIC simulation results as a guidance
for developing more accurate 2D radial-axial PIC models.
For example, providing an accurate 2D profile of the electron mobility,
an anomalous collision frequency for electron transport would be
directly evaluated,
% through the relation
% \begin{equation}
%     \mu'_{\perp} =
%     - \dfrac{\dfrac{e}{m \nu_a}}{1 + \dfrac{\omega^2_{ce}}{\nu_a^2}},
% \end{equation}
which would save a lot of computational cost and could eventually leads
to a successful engineering model for guiding Hall thruster design.

% \begin{figure}[ht]
% \centering
% \includegraphics[width=0.4\textwidth]{Bhalf.png}
% \caption{
% The correlation $\left< n_e E_y \right>_{B_{\textrm{half}}}$ of a linear fitted
% result in between
% the case Strong-B and Weak-B.
% }
% \label{fig:Bhalf}
% \end{figure}

Prior to this, however,
the 3D PIC model must be improved such that the simulated conditions
match those in the experiments.
On the one hand,
more realistic simulation setups and physical models are needed,
such as ceramic surface boundary condition with secondary electron emissions,
more accurate solution of the neutral gas flow,
either bigger channel and plume domain size or better ambient boundary condition
to weaken the non-physical effects,
and of course applying more number of macro-particles
under smaller cell size and timestep.
The simulations results must be verified by comparing to
experimental diagnoses, and some models may have to be further improved.
On the other hand,
it would be impossible to carry out fully 3D PIC simulations
for every type of Hall thrusters (with different structures)
under every operating conditions (with different voltages, flow rates, magnetic field
strengths, propellants)
due to the large computational cost and time consumed.
Thus, one way around would be to pick characteristic thruster types and operating conditions to conduct
a feasible number of 3D simulations as trustable provided data points.
Then, these data could be applied to fit analytical functions or train machine learning
models to interpolate and extrapolate throughout the entire parameter space.

% Take the data obtained in this paper for example,
% a linear fit can be done for the correlation term $\left< n_e E_y \right>$
% using the Strong-B case and the Weak-B case.
% Predictions of $\left< n_e E_y \right>$ under intermediate magnetic field strengths
% could be predicted,
% such as when the magnetic field is right in between of the two cases,
% labeled by $B_{\textrm{half}}$,
% the correlation $\left< n_e E_y \right>_{B_{\textrm{half}}}$ could be approximated by
% \begin{equation}
%     \left< n_e E_y \right>_{B_{\textrm{half}}} = 
%     \left< n_e E_y \right>_m
%     \dfrac{\left< n_e E_y \right>^*_{B1} + \left< n_e E_y \right>^*_{B2}}{2},
% \end{equation}
% where the subscript $B1$ denotes the Weak-B case,
% $B2$ denotes the Strong-B case,
% the superscript $*$ denote a normalization
% such that the maximum value of $\left< n_e E_y \right>^*$ equals one,
% and $\left< n_e E_y \right>_m$ is a scalar
% equaling the average between the maximum values of
% $\left< n_e E_y \right>_{B1}$
% and
% $\left< n_e E_y \right>_{B2}$.
% The result is shown in
% Fig.\ref{fig:Bhalf},
% from which an intermediate profile can be seen
% that the electron transport path is moving from the downstream
% region of the Weak-B case to the upper-radial region
% near the exit of the Strong-B case.

The main conclusions of this paper are summarized as follows:

(1) The commonly used analytical magnetic field model that considers only the radial component is found to yield electron transport characteristics that differ significantly from those obtained under a more realistic model incorporating both radial and axial components in 3D PIC simulations.

(2) Practical variations in the radial magnetic field strength lead to highly asymmetric electron transport, with enhanced transport occurring in regions of lower magnetic field strength.

(3) The effective electron mobility, $\mu'_{\text{eff}}$, computed from the correlation between electron density and the azimuthal electric field, generally agrees with the mobility $\mu_{ez}$ directly extracted from PIC simulations.

(4) Finally, the 2D profiles of electron transport ($\langle n_e E_y \rangle$ or $\mu'_{\text{eff}}$) derived from the 3D simulations suggest a pathway for developing more accurate 2D radial-axial PIC models by leveraging 3D simulation results as a guide.

% \section*{Supplementary Material}

\section*{Acknowledgment}

The authors acknowledge the support from National Natural Science Foundation of China (Grant No. 5247120164).

\section*{Data Availability}

The data that support the findings of this study is available from
the corresponding author upon reasonable request.

\renewcommand{\bibsection}{\section*{References}}
\bibliographystyle{unsrturl}
\bibliography{reference}

\end{document}